\documentclass[12pt]{article}
\usepackage[maxauthors=100,articletitle=true]{achemso}
\usepackage[version=4]{mhchem}
\usepackage[scale=0.80]{geometry}
\usepackage[utf8]{inputenc}
\usepackage[english]{babel}
\usepackage{amsmath, amssymb, amsfonts}
\usepackage{multirow}
\usepackage{titlesec}
\usepackage{caption}
\usepackage{ragged2e}
\usepackage[colorlinks=true,linkcolor=blue,citecolor=blue,urlcolor=blue]{hyperref}
\usepackage{array}
\usepackage{booktabs}
\usepackage{tabularx}
\usepackage{siunitx}

\usepackage{graphicx}
\usepackage{float}
\usepackage{setspace}
\usepackage[numbib,nottoc]{tocbibind}
\usepackage{color, colortbl}
\usepackage{enumerate}
\usepackage[table]{xcolor}
\usepackage{natbib}
\usepackage{authblk} % For multiple authors and affiliations
\usepackage{indentfirst}

\usepackage{multirow}%
\usepackage{amsmath,amssymb,amsfonts}%
\usepackage{amsthm}%
\usepackage{mathrsfs}%
\usepackage[title]{appendix}%
\usepackage{xcolor}%
\usepackage{textcomp}%
\usepackage{manyfoot}%
\usepackage{booktabs}%
\usepackage{algorithm}%
\usepackage{algorithmicx}%
\usepackage{algpseudocode}%
\usepackage{listings}%

\makeatletter
\renewcommand{\maketitle}{
\begin{center}
{\LARGE \textbf{\@title}} \\[1em]   % Title font size and style
{\large \@author} \\[1em]           % Author font size
{\small}                     % Date font size
\end{center}
}
\makeatother

\renewcommand{\arraystretch}{1.5}

\titleformat{\section}{\large\bfseries}{\thesection}{1em}{}
\titleformat{\subsection}{\normalsize\bfseries}{\thesubsection}{1em}{}

\title{An Active Learning Interatomic Potential For Defect-Engineered CoCrFeMnNi High-Entropy Alloy}

\author[1,$\dagger$]{Manish Sahoo}
\author[2,3,$\dagger$]{Akash Deshmukh}
\author[2]{Yash Kokane}
\author[4]{Jayaprakash H M}
\author[2,*]{Raghavan Ranganathan}

\affil[1]{\small Department of Advanced Materials Science, GSFS, University of Tokyo, Kashiwa, Japan}
\affil[2]{\small Department of Materials Engineering, Indian Institute of Technology Gandhinagar, Gujarat, 382355, India}
\affil[3]{\small Herbert Gleiter International Institute, Liaoning Academy of Materials, Shenyang-110167, China.}
\affil[4]{\small Regional Institute of Education, Manasagangothri, Mysuru, Karnataka, India}

\begin{document}
\emergencystretch 3em
\maketitle

\begin{center}
    $\dagger$These authors contributed equally to the work\\
    *Corresponding author; e-mail: rraghav@iitgn.ac.in\\
\end{center}

\begin{abstract}

High-entropy alloys (HEAs) exhibit exceptional properties arising from a combination of thermodynamic, kinetic and structural factors and have found applications in numerous fields such as aerospace, energy, chemical industries, hydrogen storage, and ocean engineering. However, a large compositional space remains to be explored. Unlike conventional approaches, computational methods have shown accelerated discovery of novel alloys in a short time. However, the lack of interatomic potentials have posed a challenge in discovering new alloy compositions and property measurements. In the present work, we have developed a Moment Tensor Potential (MTP) trained by Machine Learning based approach using the BFGS unconstrained optimization algorithm for the CoCrFeMnNi High-entropy alloy. Our training set consists of various defects induced configurations such as vacancies, dislocations and stacking-faults. An active learning scheme to re-train the potential was undertaken to dynamically to add training data upon encountering extrapolative configurations during non-equilibrium simulations. A thorough investigation of the error metrics, equation of state, uniaxial tensile deformation, nano-indentation and solid-liquid interface stability for this alloy was carried out, and it is seen that the MTP potential outperforms the popular Modified Embedded Atom Method (MEAM) potential on physical properties prediction. The accuracy and high computational speed are discussed using scaling performance. The potential is prepared for public use by embedding it into the Large-scale Atomic/Molecular Massively Parallel Simulator (LAMMPS) code.

\end{abstract}

\section{Introduction}

Advances in materials have led to the design and development of novel structural and functional materials for a variety of applications \cite{krishna2024comprehensive,luo2024high}. The conventional alloy design strategy was always centered on one or two principal elements dominating the alloying system and adding small amount of other elements to modify the properties  \cite{yu2024recent,zhou2023research,nene2024metallurgical}. This design strategy was found to reach its limits to develop novel alloys \cite{jafary2019applying}. These constraints motivated researchers to invent new alloys that led to the advent of high-entropy alloys (HEAs) \cite{jafary2019applying}. The concept of HEAs was understood while studying the intermetallics \cite{yu2024recent, yang2022chemically,pasini2023boron} and bulk metallic glasses \cite{yu2024recent,mahbooba2019additive,gao2022recent}. In 2004, this concept was published by Yeh et al. \cite{yeh2004nanostructured} and Cantor et al. \cite{cantor2004microstructural} independently. HEAs possess excellent strength and hardness \cite{zhou2023research,krishna2024comprehensive,xiang2024review}, high fracture and fatigue resistance \cite{krishna2024comprehensive,xiang2024review}, excellent tribological properties \cite{zhou2023research,xiang2024review} and high temperature stability \cite{zhou2023research,krishna2024comprehensive}. They also have extraordinary radiation\cite{zhou2023research}, oxidation,\cite{zhou2023research} and corrosion \cite{zhou2023research,krishna2024comprehensive} resistance, superparamagnetic \cite{xiang2024review} and superconductivity properties \cite{xiang2024review}. They find applications in wider fields including but not limited to national defense \cite{zhou2023research}, aerospace\cite{zhou2023research,xiang2024review,li2021recent}, medical equipment\cite{zhou2023research}, mining machinery\cite{zhou2023research}, nuclear industry\cite{zhou2023research,xiang2024review}, jet turbine blade manufacturing \cite{xiang2024review}, liquid fuel equipment \cite{li2021recent}, biomimetic materials \cite{li2021recent}, magnetic resonance imaging (MRI) scanners \cite{li2021recent}, hydrogen storage \cite{luo2024high,ding2025integral} and catalysis \cite{li2021recent}. The vast configurational space for alloy design in HEAs arising from the optimization of the number of elements and the composition has verily fascinated scientists in the pursuit for finding promising systems for enhanced properties \cite{zhou2023research}.

The origin of the extraordinary properties of these alloys is rooted in the four core effects proposed by Yeh et al. \cite{yeh2004nanostructured}. Briefly, these are (1) the thermodynamic high-entropy effect; (2) the kinetic sluggish diffusion effect; (3) structural effect from severe lattice distortion; and (4) a cocktail effect arising from a mixture of properties \cite{yeh2004nanostructured}. The primary motivation in studying HEAs were based on the fact that the configurational entropy favors a single-phase solid solution (SS) against intermetallic (IM) compounds \cite{bosi2023empirical}. The prominent single-phase SS formed in the HEAs are BCC, FCC or HCP \cite{alam2023revisiting,qiu2017lightweight,rogal2020design}. However, reports have shown that it is difficult to achieve the entropy stabilized single-phase HEA with strength-ductility synergy for industrial applications \cite{li2017trip,jain2025dual}. Interestingly, the presence of dual-phase or multiphase has shown excellent strength-ductility synergy compared to single-phase SS\cite{jain2025dual}. These phases can be BCC + HCP, BCC + FCC, FCC + HCP, and BCC + FCC + HCP) \cite{alam2023revisiting, kube2019phase}. In some cases, secondary phases such as IM and amorphous phases have also been reported \cite{alam2023revisiting, kube2019phase,hossain2022microstructure}.
The IM phases are found to lower the ductility and lead to difficulty in HEA processing \cite{alam2023revisiting,qiu2017lightweight}. Therefore, the trend is shifting to achieve the dual or multiphases by developing the second generation HEAs with non-equiatomic ratios exhibiting the excellent combination of properties 
\cite{hossain2022microstructure,liu2023tribology}. These studies provided the space to relax the restriction on the formation of single phase SS in HEA \cite{li2017trip,liu2023tribology}. Further, challenges were to predict the SS, crystal structure, phase stability \cite{qiu2017lightweight, kube2019phase, yang2020revisit, alam2023revisiting}. Combinatorial theory suggested a wide compositional space at the centre of the phase diagram for ternary alloys \cite{singh2023phase, kube2019phase}. According to this theory, approximately $10^8$ types of HEAs can be developed from the 64 elements in the periodic table \cite{singh2023phase, kube2019phase}. This space is five orders of magnitude more than the presently discovered $<10^3$  quinary HEAs \cite{singh2023phase, kube2019phase}.\newline

Experimentally, it is impossible to explore such a wide spectrum of compositions by trial-and-error methods \cite{sun2024machine}. Though empirical models such as phase formation rules \cite{qiu2017lightweight, kube2019phase, yang2020revisit, alam2023revisiting} and ductility criteria \cite{mak2021ductility} are available for guiding new compositions, they are highly restricted due to the cosmic variety of compositions \cite{sun2024machine}. Besides, it is insignificant to design empirical atomic interaction models due to the large number of chemical interactions that will increase exponentially with higher-order of alloys \cite{sun2024machine}. Though, computational approaches such as phase diagram calculation (CALPHAD) \cite{sun2024machine,martin2022heaps,qiao2021focused,jiang2024accelerating, silva2024computational,kumar2022novel, singh2023phase,odetola2024exploring}, Density Functional Theory
(DFT) \cite{sun2024machine,martin2022heaps, qiao2021focused, singh2023phase, jiang2024accelerating, odetola2024exploring}, and molecular dynamics (MD) simulations \cite{sun2024machine,martin2022heaps, qiao2021focused, jiang2024accelerating, singh2023phase}  have been successfully used to explore HEAs, the high computational cost, time consumption, and uncertainty of these methods have severely hindered their application for enabling quick property prediction and screening of HEAs. The CALPHAD method is useful in predicting complex HEAs by extrapolating the thermodynamic properties of binary and ternary alloys \cite{zhang2022calphad, odetola2024exploring}. Although the CALPHAD model can potentially help design novel HEAs and select the best compositions, it observes some limitations \cite{zhang2022calphad,odetola2024exploring}. The reliance on empirical data and the inconsistencies arising from thermodynamic calculations limit their predictive capacity \cite{odetola2024exploring, zhang2022calphad}. Similarly, CALPHAD had shown lower precision in predicting IM phases such as Laves (C14, C15, C36 and PuNi$_3$), A15, chi($\chi$), mu($\mu$), and sigma ($\sigma$) \cite{silva2024computational}. Alternatively, DFT-based calculations, such as first-principles and ab initio molecular dynamics (AIMD), have emerged as valuable tools to inspect thermophysical and mechanical properties \cite{sun2024machine,martin2022heaps, qiao2021focused, singh2023phase, jiang2024accelerating, odetola2024exploring}. Although DFT simulations have meticulously advanced our understanding of the phase, structure, and various properties of HEAs, these calculations required small simulation cell sizes owing to the computational complexity and relatively long simulation times, resulting in high computational cost \cite{sun2024machine,martin2022heaps, qiao2021focused, singh2023phase, jiang2024accelerating, odetola2024exploring,rybin2024thermophysical}. Unlike DFT, MD enables the simulations of large cell size over extended period with significant numerical precision \cite{rybin2024thermophysical,sun2024machine,martin2022heaps,qiao2021focused, jiang2024accelerating, singh2023phase}. However, this accuracy depends on the quality of the interatomic potentials \cite{rybin2024thermophysical,sun2024machine,martin2022heaps, qiao2021focused, jiang2024accelerating, singh2023phase}. In the recent past, remarkable progress has been observed in developing the machine learning interatomic potentials (MLIPs) \cite{rybin2024thermophysical, deringer2019machine}. MLIPs have demonstrated promise for MD simulations with near ab initio accuracy, on time and length scales comparable to traditional interatomic potentials \cite{rybin2024thermophysical, deringer2019machine, wu2024machine, kistanov2025unified,rybin2024moment,liu2024first,klimanova2025accelerating,ito2024machine,stark2023machine,podryabinkin2023mlip,pandey2022machine,zuo2020performance,jafary2019applying,song2025machine,wang2025harnessing,wu2024machine}.

Among available ML-based potentials, the Moment Tensor Potential (MTP) has shown excellent balance between computational cost and accuracy \cite{rybin2024moment,rybin2024thermophysical,klimanova2025accelerating,ito2024machine,mtp,novikov2020mlip}. Pandey et al. \cite{pandey2022machine} successfully developed the MTP based MLIP for the quaternary MoNbTaW and quinary MoNbTaTiW HEAs. They studied the mechanical properties of both alloys validating the experimental results \cite{pandey2022machine}. Similarly, Zadeh et al. \cite{jafary2019applying} developed MLIP to investigate the local lattice distortion effect (LLD) in CoFeNi. Using MD simulations, they investigated the independent effect of static, dynamic,thermal expansion and chemical short-range order (CSRO) on the LLD \cite{jafary2019applying}. Yin et al. \cite{yin2021atomistic} developed MTP to investigate the influence of CSROs on dislocation mobility in the refractory HEA MoNbTaW. Hodapp and Shapeev \cite{hodapp2021machine} investigated the effect of varying compositions on the unstable stacking fault energies (SFEs) of Mo-Nb-Ta alloys using MTP. These results were in excellent agreement with the DFT calculations \cite{hodapp2021machine}. Active learning (AL) was also demonstrated to reduce the training data size, which presents a feasible way to frameworks for guiding the high-throughput screening of HEAs \cite{hodapp2021machine}. Based on MTP-assisted MD simulations, Gubaev et al. \cite{gubaev2021finite} reported that the variation in Ta is responsible for the phase change in TiZrHfTa, which influences the elastic properties of the alloy. 

Inspired by the aforementioned approaches and challenges in developing MLIPs for HEAs, the present work demonstrates the development of a Moment Tensor Potential (MTP) MLIP for the equiatomic CoCrFeMnNi HEA, also called the Cantor alloy. Owing to its equiatomic composition, this alloy has contributed to the extraordinary tensile strength, ductility and fracture toughness resulting in long service life and durability of the equipment \cite{cantor2004microstructural, konovalov2024evolution,yang2020competing,yang2021influence,zhang2023comparative,hu2022microstructure,zhang2023effect}. This alloy also exhibits significant wear and corrosion resistance \cite{cantor2004microstructural, konovalov2024evolution,yang2020competing,yang2021influence,zhang2023comparative,hu2022microstructure,zhang2023effect}. 
In this work, we have trained the MTP MLIP for the equiatomic CoCrFeMnNi HEA using a large dataset of DFT configurations generated using the Special Quasi-random Structure (SQS) method, with perfect as well as defect-engineered frames over large temperature ranges. Active Learning was incorporated to incorporate new ``extrapolative" configurations into the training data, significantly augmenting the MLIP's ability to handle a wide range of atomic environments. The model was validated with respect to DFT data through error metrics like force, energy and stress accuracy. The model was also compared with a popular parametrization of the MEAM (Modified Embedded Atom Model) potential for the Cantor alloy, through simulations for determining lattice parameter, solid-liquid phase transition and bulk modulus. We have also recently employed this active learning MTP for elucidating the complex role of defects on the high-frequency viscoelastic damping of the Cantor alloy \cite{deshmukh2025viscoelastic}. Finally, a brief discussion of the scaling performance of the MTP potential is presented. 

\section{Computational Methods}

\begin{figure}[H]
    \centering
\includegraphics[width=0.95\linewidth,keepaspectratio]{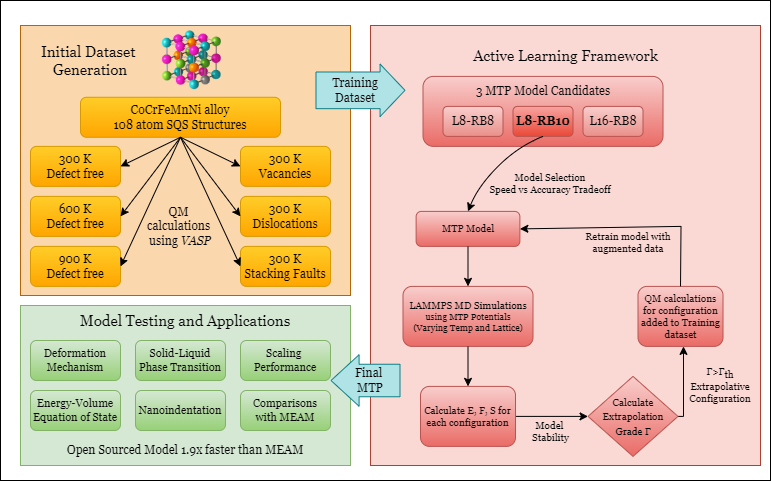}
\caption{Schematic overview of the MLIP workflow.}

    \label{fig:TableOfContents}
\end{figure}

A schematic overview of the complete workflow for developing the machine learning interatomic potential (MLIP) is presented in Fig(\ref{fig:TableOfContents}). The procedure began with the construction of a diverse initial training set for the CoCrFeMnNi HEA. To capture the alloy's chemical complexity, special quasi-random structures (SQS) were generated using the \texttt{mcsqs} (Monte Carlo Special Quasirandom Structure) code from the Alloy Theoretical Automated Toolkit (ATAT) \cite{atat2}, to obtain  representative HEA atomic configurations. This technique samples atomic configurations that reproduce the statistical  correlation functions of an ideal random alloy for a chosen set of neighbor shells, within the constraints of a small (periodic) simulation cell sizes that are possible at the level of DFT calculations. An important contribution of the current work is to ensure that the trained MLIP can capture a realistic description of defects including vacancies, stacking faults and dislocations under a variety of active learning environments (such as various states of deformation and various temperatures), thereby lending a high degree of physical relevance and applicability. High-fidelity reference data—energies, forces, and stresses—for all initial configurations were calculated using DFT, that served as the initial dataset was used to train a baseline Moment Tensor Potential (MTP). Subsequently, a rigorous active learning (AL) cycle was initiated to automatically refine and expand the potential's applicability. This iterative process involved using the current MTP within the Large-scale Atomic/Molecular Massively Parallel Simulator (LAMMPS) package \cite{plimpton1995fast} for structural optimization tasks while continuously monitoring for extrapolation. Any configuration identified as novel or ``extrapolative" was automatically selected for a new DFT calculation. The resulting data was then added to the training set, and the MTP was retrained. This cycle of \textit{simulation $\rightarrow$ extrapolation check $\rightarrow$ DFT calculation $\rightarrow$ retraining} was repeated until the potential achieved robust stability across all required simulation conditions.

As part of our model selection process, a comparative study was performed, primarily evaluating the performance of MTP potentials with different hyperparameters. The final, fully validated MTP was then deployed for large-scale MD simulations to investigate the alloy's properties, such as deformation mechanisms, phase transitions, equilibrium lattice parameters, along with an analysis of the trained model's accuracy and scaling performance. 

\subsection{Moment Tensor Potentials (MTP)}

In this work, the moment tensor potentials (MTPs) with active learning are employed as
implemented in the MLIP package.\cite{novikov2020mlip,mtp}
Within the MTP framework, the total potential energy $E^{\text{MTP}}$ of an atomic
configuration is expressed as a sum of local atomic contributions,
\begin{equation}
    E^{\text{MTP}} = \sum_{i=1}^{N} V(n_i),
\end{equation}
where $i$ indexes atoms, $n_i$ denotes the local atomic neighborhood of atom $i$ within a
cutoff radius $R_{\text{cut}}$, and $V(n_i)$ is the energy associated with that atomic
environment.

The local energy $V(n_i)$ is written as a linear combination of invariant basis functions,
\begin{equation}
    V(n_i) = \sum_{\alpha} \xi_{\alpha}\, B_{\alpha}(n_i),
\end{equation}
where $\xi_{\alpha}$ are linear fitting coefficients and $B_{\alpha}(n_i)$ are symmetry-invariant
descriptors of the environment of atom $i$.

These basis functions are constructed from so-called moment tensors. For each environment
$n_i$, the moment tensors $M_{\mu\nu}(n_i)$ are defined as
\begin{equation}
    M_{\mu\nu}(n_i) = \sum_{j \in n_i}
    f_{\mu}\bigl(|\mathbf{r}_{ij}|\bigr)\,
    \bigl(\mathbf{r}_{ij}^{\otimes \nu}\bigr),
\end{equation}
where $\mathbf{r}_{ij}$ is the vector from atom $i$ to its neighbor $j$, $f_{\mu}$ are radial
basis functions, and $\mathbf{r}_{ij}^{\otimes \nu}$ denotes the $\nu$-fold tensor product of
$\mathbf{r}_{ij}$. The scalar basis functions $B_{\alpha}(n_i)$ entering Eq.~(2) are then defined
as all possible contractions of one or more moment tensors $M_{\mu\nu}(n_i)$, which yields
a systematically improvable set of rotationally, permutationally, and translationally invariant
descriptors of the local atomic environment.
More details on the complete description of the MTP can be found in the previous reports \cite{gubaev2019accelerating,novikov2020mlip,liu2023101018, podryabinkin2023mlip,rybin2024thermophysical,rybin2024moment, klimanova2025accelerating}. 

\subsection{Construction of Training Dataset}

It is anticipated that a robust interatomic potential for HEAs must accurately capture the energetics of diverse local atomic environments, including perfect crystal structures, defective configurations, and thermally activated states.  Our MLIP achieves this by training on a comprehensive data set employing DFT calculations spanning multiple temperatures and including common crystal defects. The construction of the data set involved steps as elaborated below: 

\subsubsection{Special Quasirandom Structure (SQS) Generation}

A fundamental challenge in the first principles modeling of high-entropy alloys is the faithful representation of their inherent chemical disorder within the constraints of a computationally tractable, periodic simulation cell. A simple random placement of atoms in a small supercell often leads to spurious short-range ordering and fails to capture the average properties of the macroscopic material. To overcome this, we employed the Special Quasi-random Structure (SQS) method, a robust and widely-used technique for designing supercells that accurately mimic the statistical correlation functions of a truly random alloy \cite{jiang2004first, shin2006thermodynamic, liu2009first}. The guiding principle of the SQS approach is to perform a systematic search for a specific atomic arrangement where the correlation functions, $\prod_{k,m}(\mathrm{SQS})$, for pairs, triplets, and other small clusters match those of a perfectly random alloy, $\prod_{k,m}(\mathcal{R})$, as closely as possible \cite{yang2021ab}.

The SQS configurations for this study were generated using the \texttt{mcsqs} code, a powerful tool within the Alloy Theoretical Automated Toolkit (ATAT) package \cite{atat2, van2013efficient}. Our target was a 108-atom cubic supercell, constructed via a \(3\times3\times3\) expansion of a conventional 4-atom FCC unit cell. The initial lattice constant was set to $a=3.518\,\mathrm{\AA}$, a value derived from preliminary DFT convergence tests that established the optimal k-point meshes and plane-wave energy cutoffs for our system. To closely approximate the ideal equiatomic ratio within the 108-atom cell, the supercell was populated with a specific composition of Co$_{22}$Cr$_{22}$Mn$_{22}$Fe$_{21}$Ni$_{21}$. The search for the optimal atomic configuration was computationally intensive, involving an objective function minimization over approximately $10^9$ randomly generated structures. This objective function was carefully defined to prioritize the statistical match for the most physically significant atomic interactions. Specifically, the fit for the first, second, and third nearest-neighbor shells was weighted, with the closest shells receiving the highest importance to ensure that the local chemical environments are as random as possible. From the top-ranked candidates produced by this exhaustive search, the single SQS configuration with the lowest objective function score—representing the minimum possible deviation from perfect randomness—was selected for all subsequent DFT calculations.

\subsubsection{Defect Engineering}
To introduce the variations in the training set, three types of defects namely, vacancies, dislocations, and stacking fault were created in the SQS structures. These defects probe the potential’s ability to handle local undercoordination, large lattice distortions, and planar defects, respectively.

\begin{enumerate}
    \item \textbf{Vacancy:} Three atoms were removed from the bulk region of the SQS to create vacancy sites (periodic boundary conditions ensured no spurious interactions). This yields a system size of 105 atoms with the composition balanced as closely as possible among Fe, Mn, Co, Cr, and Ni atoms.
    \item \textbf{Dislocation:} A half-plane of atoms was removed along the \{111\} plane to be consistent with \(\langle110\rangle\)\{111\} primary slip system for FCC lattice. This yields a system size of 103 atoms.
    \item \textbf{Stacking Fault:} A single plane of atoms was shifted by one-third of the interatomic spacing on the \{111\} plane.
\end{enumerate}

\subsubsection{Quantum Mechanical Calculations}
\emph{Ab initio} molecular dynamics (AIMD) calculations were performed for each of the aforementioned structure in the DFT framework of Vienna ab initio simulation package (VASP)\cite{kresse1996efficient,kresse1996efficiency}
The  exchange and correlation effects were observed with the 
Perdew–Burke–Ernzerhof (PBE) exchange-correlation function\cite{perdew1996generalized} under the generalized gradient approximation (GGA) \cite{perdew1992atoms}. For energy calculations, A \(\gamma\)-centered \textit{k}-mesh was chosen with spacing commensurate to the supercell size (e.g., approximately $2\pi \times 0.03 \mathrm{\AA}^{-1}$ were used to sample the Brillouin zones. The total energies obtained by relaxation calculations were converged to better than $10^{-5} \mathrm{eV}$, and ionic relaxations were proceeded until forces were below $0.02 \mathrm{eV/\AA}$.
A plane-wave energy cutoff of \SI{400}{eV} is used for all calculations. For the defect-free SQS, AIMD simulations were performed at \(300\,\mathrm{K}\), \(600\,\mathrm{K}\), and \(900\,\mathrm{K}\) to capture thermal effects and diverse atomic configurations. Configurations were periodically extracted from the equilibrated portions of these trajectories for DFT single-point evaluations. The data set consists of 42 configurations each for \SI{300}{K}, \SI{600}{K}, and \SI{900}{K}. Next to this, defect-engineered structures (vacancy, dislocation, and stacking fault) were simulated at \(300\,\mathrm{K}\), reflecting typical room-temperature conditions for capturing defect energetics without additional thermal disorder. The data set was further increased by considering the 42 configurations for each vacancy and dislocation. In addition, 30 configurations for stacking faults were considered. In total, the initial data set consists of 240 configurations for training the MTP. It is to be noted that this training set is only for the initial model. Subsequent improvements to the model were added through active learning. The initial data set consisting of $conf_{k}$ configurations where k=1,..., 240. For each configuration energy, k,  $E^{DFT}(conf)$, force $f^{DFT}(conf)$ and stress tensor $\sigma^{DFT}(conf)$ were calculated. 
The complete data set was divided into training set and test set with a ratio of 80:20 as illustrated in Table \ref{tab:overallDFTdata}. This extensive dataset ensures that the MTP is exposed to relevant physical environments, from perfect crystal structures under varying thermal loads to local defect distortions at \(300\,\mathrm{K}\).      

\subsubsection{Conversion of DFT dataset to MLIP-compatible format}
It is important to note that the output files generated using DFT calculations can not be used directly to initiate the MTP training. We have converted the DFT data output consisting of atomic coordinates, energies, forces and stresses into the CFG file format that is compatible with the MLIP package\cite{novikov2020mlip}. In this way, a final dataset required for training the MTP was created.

%All configurations were converged using DFT with a generalized-gradient approximation (GGA) functional (e.g., PBE) as implemented in a plane-wave code such as VASP or Quantum ESPRESSO:
%\begin{itemize}
    %\item \textbf{Plane-Wave Cutoff:} Set to a sufficiently large value (e.g., \(\geq 400\,\mathrm{eV}\)) to ensure accurate energies and forces.
    %\item \textbf{k-Point Sampling:} A \(\gamma\)-centered \textit{k}-mesh was chosen with spacing commensurate to the supercell size (e.g., approximately \(2\pi \times 0.03\,\mathrm{\AA}^{-1}\)).
    %\item \textbf{Convergence Criteria:} Total energies converged to better than \(10^{-5}\,\mathrm{eV}\), and ionic relaxations proceeded until forces were below \(0.02\,\mathrm{eV/\AA}\).
%\end{itemize}

%---------------------------------------
% Latex code for the summary table
%---------------------------------------
\begin{table}[H]
    \centering
    %\caption{Overview of DFT configurations in the final dataset. All defect structures (dislocation, stacking fault, vacancy) were simulated at $300\,\mathrm{K}$.}
    \caption{Defect-free and defect-engineered DFT configurations in the dataset. Defect-engineered structures were simulated at $300\,\mathrm{K}$.}
    \label{tab:overallDFTdata}
    \arrayrulecolor{black} % Set table borders to black
    
    \begin{tabular}{|l|c|c|c|c|S[table-format=1.2]|}
        \toprule
        \hline
        \textbf{Category} & \textbf{Temperature} & \textbf{Training} & \textbf{Test} & \textbf{Total} \\
        \midrule
        \hline
        Defect-free   & 300\,K & 36 & 6  & 42 \\
        Defect-free   & 600\,K & 36 & 6  & 42 \\
        Defect-free   & 900\,K & 36 & 6  & 42 \\
        \midrule
        Dislocation     & 300\,K & 36 & 6  & 42 \\
        Stacking Fault  & 300\,K & 20 & 10 & 30 \\
        Vacancy         & 300\,K & 36 & 6  & 42 \\
        \midrule
        \hline
        \textbf{Total}  & --     & 200 & 40 & 240 \\
        \bottomrule
        \hline
    \end{tabular}
\end{table}

\subsection{Model Building and Its Stabilization}

The feasibility of the MLIP used for the study depends on its ability to represent the Potential Energy Surface for the multi-component alloy without being excessively complex. For Moment Tensor Potentials (MTPs), the architecture is primarily defined by a set of hyperparameters that dictate the flexibility and descriptive power of the potential \cite{mtp}. The goal is to find an optimal balance between predictive accuracy and generalizability, while maintaining computational tractability \cite{zuo2020performance}. The most critical of these hyperparameters are the potential's ``level" and the size of its radial basis set.

\begin{enumerate}
    \item \textbf{Level}: This integer hyperparameter controls the maximum rank of the moment tensors used to construct the basis functions, which in turn governs the angular complexity of the potential. A higher level allows the MTP to capture more sophisticated many-body interactions. For instance, a low-level potential might accurately describe simple bond angles (three-body terms), whereas a higher-level potential can represent more complex geometries like torsional angles and the intricate local environments found near defects or in amorphous structures (four-body and higher-order terms) \cite{gubaev2019accelerating}. However, increasing the level leads to a rapid, combinatorial increase in the number of parameters to be fitted. This not only increases the computational cost of training and evaluation but also raises the risk of overfitting if the training data is not sufficiently diverse to constrain the additional parameters.

    \item \textbf{Radial Basis Size (\(N_\mathrm{rb}\))}: This parameter determines the number of radial basis functions used to describe how interatomic interactions change with distance. In the MLIP package implementation of MTP, Chebyshev polynomials are used as the basis functions due to their numerical stability and flexibility. A larger radial basis size (\(N_\mathrm{rb}\)) provides the model with greater flexibility to resolve fine features in the radial distribution of neighboring atoms and precisely model the shape of the two-body potential. While this can improve accuracy, an overly large basis can also lead to overfitting by fitting noise in the training data, resulting in unphysical oscillations in the potential and poor performance on unseen configurations \cite{novikov2020mlip}.
\end{enumerate}

To identify an architecture suitable for the chemically complex CoCrFeMnNi HEA, we systematically explored three candidate MTPs that probe different aspects of this trade-off between complexity and robustness:
\begin{itemize}
    \item \textbf{Level 8, Radial Basis Size 8 (L8-RB8)}: A baseline configuration that has proven effective for simpler metallic systems.
    \item \textbf{Level 8, Radial Basis Size 10 (L8-RB10)}: This model investigates the benefit of increased radial flexibility at a fixed, moderate level of angular complexity.
    \item \textbf{Level 16, Radial Basis Size 8 (L16-RB8)}: This configuration tests whether a significant increase in angular complexity is necessary to capture the physics of the HEA.
\end{itemize}

A systematic comparison of the three candidate models was performed. This comparison was guided by a central question regarding the most efficient path to improving the potential's fidelity. Previous work has suggested that Level 8 MTPs can be sufficient for a variety of systems \cite{meng2025small, leimeroth2025machine}. The jump from Level 8 to Level 16 markedly increases the number of angular moment channels, offering a more complex description of many-body geometries, but at a significant computational cost. Conversely, increasing the radial basis size (e.g., from 8 to 10) directly augments the resolution of distance-dependent features, which can also enhance accuracy, often at a lower computational penalty. Thus, the performance of these candidate potentials was evaluated to select a final model that provides the best compromise between accuracy, stability in molecular dynamics simulations, and computational cost. The detailed comparison and justification for our final model choice are presented in Section \ref{sec:model_selection}.

\subsubsection{Training Process and Loss Function}

The initial training of the MTP is a ``passive learning" stage, where the potential's parameters are optimized to reproduce a fixed, pre-existing set of DFT reference data. This process aims to find the optimal set of linear MTP parameters, denoted collectively by the vector $\theta$, that minimizes a scalar loss function, $\mathcal{L}(\theta)$. The loss function quantifies the discrepancy between the MTP's predictions and the DFT ground truth for a training set of $K$ configurations.

The loss function is defined as a weighted sum of the squared errors in the predicted energies ($E$), atomic forces ($\mathbf{f}$), and virial stresses ($\boldsymbol{\sigma}$), as shown in Equation \ref{eq:loss_function}:

\begin{equation}
\label{eq:loss_function}
\mathcal{L}(\theta) = \sum_{k=1}^{K} \Biggl[ 
  w_{e} \left( E^{\text{mtp}}_{k}(\theta) - E^{\text{DFT}}_{k} \right)^2 
  + w_{f} \sum_{i=1}^{N_{k}} \left\| \mathbf{f}^{\text{mtp}}_{i,k}(\theta) - \mathbf{f}^{\text{DFT}}_{i,k} \right\|^2 
  + w_{s} \left\| \boldsymbol{\sigma}^{\text{mtp}}_{k}(\theta) - \boldsymbol{\sigma}^{\text{DFT}}_{k} \right\|^2 \Biggr] 
\end{equation}

In this equation, $N_k$ is the number of atoms in the $k^{th}$ configuration. The terms $w_e$, $w_f$, and $w_s$ are tunable, non-negative weights that control the relative importance of fitting energies, forces, and stresses, respectively. A higher weight for forces, for instance, prioritizes the accuracy of the potential energy surface's gradient, which is crucial for reliable molecular dynamics simulations. For this work, we adopted the well-established default weights of $w_{e} = 1$, $w_{f} = 0.01$, and $w_{s} = 0.001$, as implemented in the MLIP software package \cite{novikov2020mlip, kistanov2025unified}.

To determine the minimum required dataset size for a stable potential before initiating the computationally expensive active learning cycles, we performed a learning curve analysis. A baseline MTP architecture of Level 8 (L8-RB8), with a cutoff radius of $R_{cut} = 5.0$ \AA\ and a minimum radius of $R_{min} = 2.0$ \AA, was trained on systematically larger, diverse subsets of the initial data, with sizes of 50, 100, 150, and 200 configurations. This analysis, detailed in Section \ref{sec:model_selection}, informed our decision on the necessary data volume for robustly training the final candidate potentials.

\subsubsection{Active Learning for Automated Training Set Refinement}
To ensure the MTP is robust across a wide range of atomic environments, we employed an active learning (AL) strategy to systematically and automatically refine the training set. This ``on-the-fly" approach iteratively identifies configurations where the potential is uncertain and adds them to the training data, thereby improving the model's reliability \cite{gubaev2019accelerating}. The core of this process is the ``extrapolation grade", which quantifies model uncertainty.

\paragraph{The Extrapolation Grade ($\gamma$).}
The MTP formalism represents the energy of any atomic environment, $\mathbf{n}_i$, as a linear combination of basis functions, $B_\alpha(\mathbf{n}_i)$. The entire training set thus defines a domain of ``known" atomic structures. Using the MaxVol algorithm \cite{gubaev2019accelerating}, a small, representative subset of the training set's basis vectors is selected to form an ``active set". The extrapolation grade, $\gamma(\mathbf{x})$, for a new, unseen atomic environment $\mathbf{x}$ is then calculated by decomposing its basis vector in terms of the basis vectors of this active set. The value of $\gamma(\mathbf{x})$ corresponds to the largest absolute coefficient in this linear decomposition \cite{podryabinkin2019accelerating}.
Conceptually, one can think of the training set as providing a library of fundamental atomic `building blocks', where each distinct local environment (e.g., a perfect FCC lattice site, a site near a vacancy) is a unique block. Each of these blocks is described by a mathematical fingerprint—its vector of basis function values, $B_\alpha(\mathbf{n}_i)$. Since the full library can be massive and redundant, the MaxVol algorithm is first used to select a small but highly representative subset of these fingerprints from the entire training set. This compact, non-redundant subset forms the `active set' and effectively defines the boundaries of the potential's knowledge \cite{gubaev2019accelerating}.

When the potential encounters a new atomic environment, $\mathbf{x}$, during an MD simulation, it attempts to describe the fingerprint of this new environment as a linear combination of the fingerprints in the active set. The extrapolation grade, $\gamma(\mathbf{x})$, is then defined as the largest absolute value of the coefficients required for this linear combination \cite{podryabinkin2019accelerating}.

A low grade ($\gamma \lesssim 2$) indicates that a configuration is well-represented by the current training data (i.e., it is an interpolation), while a high grade ($\gamma \gtrsim 10$) signals that the potential is extrapolating into an unfamiliar region of the configuration space \cite{klimanova2025accelerating}, forcing the potential to extrapolate, which may lead to unphysical results. This grade is therefore a powerful, real-time indicator of the potential's confidence.

The AL workflow was managed using the MLIP package interfaced with LAMMPS \cite{plimpton1995fast, novikov2020mlip} and proceeded according to the following iterative steps:

\textbf{Step 1: Exploration with the Current MTP.} A structural exploration task, such as a full structural relaxation or a molecular dynamics simulation, is performed using the current MTP. During this simulation, the extrapolation grade $\gamma$ is calculated at every step for every atom. A termination threshold $\gamma_{th} = 10$ is defined.
If at any point an atom's environment yields a grade $\gamma > \gamma_{th}$, the simulation is immediately halted to prevent unphysical behavior. If the simulation completes without exceeding this hard limit, it proceeds to the next step.

\textbf{Step 2: Selection of Extrapolative Configurations.} After the exploration task is complete (either by successful convergence or by termination), the entire trajectory is scanned. All atomic configurations that produced an extrapolation grade $\gamma > \gamma_{th}$ at any point are collected into a ``candidate set." This set contains all the novel environments that the current MTP found challenging.

 \textbf{Step 3: DFT Calculation and Training Set Augmentation.} To avoid adding redundant data, a representative subset of the most diverse and informative configurations is selected from the candidate set using the D-optimality criterion \cite{gubaev2019accelerating}. First-principles DFT calculations are then performed on this small, selected subset to obtain their reference energies, forces, and stresses. These newly labeled data points are then appended to the main training set.

\textbf{Step 4: Retraining the MTP.} The Moment Tensor Potential is fully retrained using the newly augmented and more comprehensive training set. This is a computationally intensive step, but it results in an improved potential that is now more accurate and robust in the regions of configurational space that were previously considered extrapolative.

This entire four-step cycle was repeated until all required simulation tasks (e.g., relaxations, heating, deformation) could be completed without flagging any new extrapolative configurations. This data-driven, iterative refinement ensures that the final potential is robust, reliable, and capable of accurately modeling the complex physics of the CoCrFeMnNi HEA.

\subsubsection{Final Model Selection}
\label{sec:model_selection}

To ensure the chosen MTP architecture provided the best balance of accuracy and computational cost, we performed a systematic comparison of the three candidate models: L8-RB8, L8-RB10, and L16-RB8. The training for each model was conducted using the MaxVol algorithm to minimize the loss function \cite{gubaev2019accelerating, novikov2020mlip}. The optimization was considered converged and halted either upon reaching a maximum of 1000 iterations or if the reduction in the validation loss remained below $10^{-4}$ for 50 consecutive steps. These criteria were chosen as a pragmatic standard to ensure robust convergence without incurring excessive computational expense.

A preliminary learning curve analysis revealed that training on subsets smaller than 200 configurations resulted in high validation errors and unstable training trajectories, indicating underfitting (high bias). We therefore used the full initial training set of 200 configurations for the definitive comparison of the three architectures. The performance characteristics of each model are summarized in Tables \ref{tab:overallDFTdata} and \ref{tab:mtp_tradeoffs}.

The results of this comparative analysis led to the following observations:
\begin{itemize}
    \item \textbf{L8-RB8}, our baseline model, showed a reasonable accuracy, but exhibited instabilities during validation using MD simulations, suggesting its radial basis was not flexible enough to capture the full complexity of the five-component interactions.
    \item \textbf{L16-RB8}, while theoretically more powerful due to its higher angular complexity, proved to be computationally demanding. Its training converged extremely slowly due to the large number of parameters (see Table \ref{tab:mtp_tradeoffs}), and the marginal improvement in accuracy over the Level 8 potentials did not justify the significant increase in computational cost for large-scale MD simulations.
    \item \textbf{L8-RB10} demonstrated the most favorable performance. It achieved a steady, monotonic decrease in both training and validation errors, yielding lower final error metrics than the L8-RB8 model. This indicates that the modest increase in radial flexibility (from 8 to 10 basis functions) was highly effective for this system, providing a superior description of the interatomic interactions without the prohibitive cost of a higher-level potential.
\end{itemize}
\begin{center}
\scriptsize
\setlength{\tabcolsep}{6pt}
\renewcommand{\arraystretch}{1.2}
\begin{tabular}{llcccccc}
\toprule
\textbf{Order-RBsize-DSsize} & \textbf{Iteration} & \textbf{Max Error} & \textbf{Avg Error} & \textbf{RMS Error} & \textbf{N\_conf} & \textbf{Sim\_Stb (steps)} \\
\midrule
8-8-50   & 1   & 0.582069  & 0.132033   & 0.175981  & 51   & 0 \\
         & 50  & 0.597468  & 0.0436068  & 0.0815622 & 105  & 0 \\
         & 100 & 0.509323  & 0.0908816  & 0.136684  & 228  & 0 \\
         & 150 & 0.533917  & 0.0541675  & 0.100096  & 399  & 0 \\
\midrule
8-8-100  & 1   & 1.12166   & 0.17794    & 0.284306  & 101  & 0 \\
         & 50  & 1.07729   & 0.107822   & 0.209704  & 150  & 0 \\
         & 100 & 1.13963   & 0.0901015  & 0.196433  & 204  & 0 \\
         & 150 & 1.02005   & 0.129674   & 0.21833   & 291  & 0 \\
\midrule
8-8-150  & 1   & 1.00392   & 0.175404   & 0.252144  & 151  & 0 \\
         & 50  & 1.19713   & 0.0939184  & 0.192938  & 200  & 0 \\
         & 100 & 1.02001   & 0.145257   & 0.231675  & 286  & 0 \\
         & 150 & 1.12872   & 0.121612   & 0.219039  & 448  & 0 \\
\midrule
8-8-200  & 1   & 2.19737   & 0.255416   & 0.453885  & 201  & 0 \\
         & 50  & 1.07485   & 0.099876   & 0.184602  & 254  & 0 \\
         & 100 & 1.04985   & 0.149764   & 0.256701  & 349  & 0 \\
         & 145 & 1.28044   & 0.1297     & 0.227283  & 528  & 0 \\
\midrule
8-10-200 & 1   & 3.68344   & 0.398851   & 0.846331  & 201  & 0 \\
         & 50  & 2.3233    & 0.236393   & 0.470713  & 249  & 200000 \\
         & 100 & 2.07803   & 0.186867   & 0.391353  & 312  & 150000 \\
         & 150 & 2.25324   & 0.143842   & 0.333817  & 447  & 50000 \\
      
\bottomrule
\end{tabular}
\normalsize
\end{center}

Based on this analysis, the \textbf{L8-RB10} architecture was identified as the optimal choice, offering the best compromise between predictive accuracy, stability, and computational efficiency. This model was therefore selected for the full active learning workflow and all subsequent property calculations presented in this work.
\begin{table}[H]
    \centering
    \caption{Evolution of Training Metrics for Different MTP Architectures and Initial Dataset Sizes.}
    \label{tab:training_evolution}
    \renewcommand{\arraystretch}{1.2}
    \sisetup{table-format=1.6, round-mode=places, round-precision=4} % Standardize number formatting
    \begin{center}

    \begin{tabular}{@{} l c c S S S c c @{}}
        \toprule
        \textbf{Model} & \textbf{DS Size} & \textbf{Iteration} & \multicolumn{3}{c}{\textbf{Energy Error (eV/atom)}} & \textbf{Configs} & \textbf{Stability} \\
        \cmidrule(lr){4-6}
        & & & {\textbf{Max}} & {\textbf{Average}} & {\textbf{RMS}} & & {\textbf{(steps)}} \\
        \midrule
        \multirow{4}{*}{L8-RB8} & \multirow{4}{*}{50} 
            & 1   & 0.582069  & 0.132033   & 0.175981  & 51   & 0 \\
            & & 50  & 0.597468  & 0.043607   & 0.081562  & 105  & 0 \\
            & & 100 & 0.509323  & 0.090882   & 0.136684  & 228  & 0 \\
            & & 150 & 0.533917  & 0.054168   & 0.100096  & 399  & 0 \\
        \midrule
        \multirow{4}{*}{L8-RB8} & \multirow{4}{*}{100}
            & 1   & 1.121660  & 0.177940   & 0.284306  & 101  & 0 \\
            & & 50  & 1.077290  & 0.107822   & 0.209704  & 150  & 0 \\
            & & 100 & 1.139630  & 0.090102   & 0.196433  & 204  & 0 \\
            & & 150 & 1.020050  & 0.129674   & 0.218330  & 291  & 0 \\
        \midrule
        \multirow{4}{*}{L8-RB8} & \multirow{4}{*}{150}
            & 1   & 1.003920  & 0.175404   & 0.252144  & 151  & 0 \\
            & & 50  & 1.197130  & 0.093918   & 0.192938  & 200  & 0 \\
            & & 100 & 1.020010  & 0.145257   & 0.231675  & 286  & 0 \\
            & & 150 & 1.128720  & 0.121612   & 0.219039  & 448  & 0 \\
        \midrule
        \multirow{4}{*}{L8-RB8} & \multirow{4}{*}{200}
            & 1   & 2.197370  & 0.255416   & 0.453885  & 201  & 0 \\
            & & 50  & 1.074850  & 0.099876   & 0.184602  & 254  & 0 \\
            & & 100 & 1.049850  & 0.149764   & 0.256701  & 349  & 0 \\
            & & 145 & 1.280440  & 0.129700   & 0.227283  & 528  & 0 \\
        \midrule
        \multirow{4}{*}{L8-RB10} & \multirow{4}{*}{200}
            & 1   & 3.683440  & 0.398851   & 0.846331  & 201  & 0 \\
            & & 50  & 2.323300  & 0.236393   & 0.470713  & 249  & 200000 \\
            & & 100 & 2.078030  & 0.186867   & 0.391353  & 312  & 150000 \\
            & & 150 & 2.253240  & 0.143842   & 0.333817  & 447  & 50000 \\
        \bottomrule
    \end{tabular}
        
    \end{center}
    \medskip % Adds a little vertical space
    \footnotesize % Use smaller font for footnotes
    \begin{flushleft} % Align footnote text to the left
        \textbf{Model:} Specifies the MTP Level and Radial Basis size (e.g., L8-RB8 is Level 8, RB size 8). \\
        \textbf{DS Size:} The initial number of configurations in the training dataset. (Dataset size) \\
        \textbf{Total Configs.:} The total number of configurations in the training set at a given iteration, including those added via active learning. \\
        \textbf{Stability (steps):} The number of molecular dynamics steps the potential could run stably before running into low-confidence configurations where the simulation breaks down due to loss of atoms. A value of N indicates the potential was not stable enough for a sustained MD run beyond N timesteps.
    \end{flushleft}
\end{table}
\begin{table}[H]
\centering
\caption{Performance trade-offs of MTP architectures}
\label{tab:mtp_tradeoffs}
\begin{tabular}{|l|c|c|c|c|}
\toprule
\textbf{Architecture} & \textbf{Parameters} & \textbf{Training Speed} & \textbf{Stability} & \textbf{MAE (\si{eV/atom})} \\
\midrule
Level 8 (RB8)           & 154  & Fast      & Moderate & 0.08 \\
\textbf{Level 8 (RB10)} & \textbf{514} & \textbf{Moderate} & \textbf{High} & \textbf{0.05} \\
Level 16 (RB8)          & 1278 & Slow      & Low      & 0.12 \\
\bottomrule
\end{tabular}
\vspace{0.5em}

\begin{minipage}{0.85\linewidth}
\footnotesize
\justifying
\textbf{MAE}: Mean Absolute Error relative to DFT reference values.\\
\textbf{Stability}: Ability to maintain physical configurations in MD simulations.
\end{minipage}
\end{table}

Based on these outcomes, we selected the L8-RB10 architecture for the full iterative workflow, balancing faster convergence with acceptable accuracy.

\subsection{L8-RB10: Active Learning and Model Stabilization}

Following the initial training phase, we employed active learning to systematically discover and incorporate new “extrapolative” configurations that were poorly represented by the existing dataset. Each round of active learning proceeded by running a classical MD simulation in LAMMPS using the current MTP, monitoring the extrapolation grade \(\gamma\) for each configuration in the trajectory. Whenever \(\gamma\) exceeded the threshold, the corresponding configuration was flagged as potentially underrepresented. Such flagged configurations were re-evaluated with DFT, and their energies, forces, and virial stresses were added to the training set. A subsequent retraining run on the augmented dataset then improved the MTP’s ability to handle these newly identified environments.\\

In our first round of active learning, we ran a 10\,ns NVT simulation at 300\,K on a 108-atom cell. We repeated the cycle MD simulation, extrapolation check, DFT evaluation, and retraining until the potential could run stably for about 1.25\,ps in this small cell without encountering any high \(\gamma\) values. Interestingly, once we transferred the same MTP to a larger cell of about 7000 atoms, it was stable for as long as 100\,ps at 300\,K. This difference reflects how a smaller cell can more easily drift into unusual configurations, making it a tougher testing ground for the potential.\\

We then pushed the method further by simulating higher temperatures (500\,K to 2000\,K) and changing the lattice parameter from 3.41\,\AA{} to 3.61\,\AA{} in steps of 0.05\,\AA{}. These conditions covered both solid and liquid states of our alloy. As in the first round, we flagged any high-\(\gamma\) snapshots, computed their DFT data, added them to the training set, and retrained. During these steps, we also ran extra checks on the model’s physical predictions. The final validation came from running large-scale (7000-atom) MD for up to 2\,ns in both NVT and NPT ensembles, as well as putting the material under tensile, compressive, and shear strains of up to 50\%. When no new snapshots triggered high \(\gamma\) values and the model remained physically reasonable throughout, we concluded that additional retraining was no longer necessary.

\subsubsection{Number of New Configurations Added}

Over the course of the active learning cycles, a certain number of previously unseen configurations were identified and appended to the original dataset. The exact total number of configurations that emerged from these extrapolative searches is 1500 over a duration of 450 active learning iterations separated into multiple stages. Regardless, the fact that the potential converged to a stable representation not only to solid phase but also solid-liquid equilibrium phases with relatively few total samples (as evidenced by the stability in simulating solid-liquid phase transition) underscores that the initial coverage of defect states, temperature variations, and lattice parameters was comprehensive and meaningful.

In total, our final model (Level 8, Radial Basis Size 10) underwent 450 iterations of active learning, spanning multiple thermodynamic states and structural perturbations. Through the combination of a broad initial dataset and iterative discovery of new configurations, the potential was refined into a form that (i) accurately reproduces DFT energies and forces for diverse defect-free and defect-containing environments, and (ii) remains stable in extended, large-scale simulations across a range of temperatures and mechanical loads.

\subsection{Model Testing and Applications}
\subsubsection{Uniaxial Tensile Deformation}
Initially, an equiatomic CoCrFeMnNi single crystal was generated in a cubic simulation cell containing \(18\times18\times18\) FCC unit cells. As each FCC cell has four lattice sites, the model comprises \(18^{3}\!\times 4 = 23{,}328\) atoms and measures \(18 a_{0}=18\times3.518\;\text{\AA}\approx63.3\;\text{\AA}\) per edge. Periodic boundary conditions were applied along all three Cartesian axes and all MD calculations were carried out with LAMMPS \cite{plimpton1995fast}. Interatomic forces were computed with two alternative potentials: (i) the MEAM parametrization of Choi et al.\,\cite{choi2018understanding}, a widely accepted potential for the Cantor alloy, and (ii) the moment-tensor potential (MTP) developed in this work; identical input decks were run once with each potential to ensure a fair comparison.\\\\
The as-built lattice was relaxed at 0 K via conjugate-gradient minimisation until the energy and force converged to $10^{-12}$ \text{eV} and $10^{-12}   \text{eV \AA}^{-1}$, respectively. Atomic velocities were then drawn from a Maxwell--Boltzmann distribution at \SI{300}{K}, and the equations of motion were integrated with the velocity--Verlet algorithm using a 1 fs timestep \cite{PhysRevE.59.3733}. A Nosé--Hoover thermostat \cite{evans_nose_hoover_1985} (damping time \(=\!0.1\) ps) and barostat (damping time \(=\!1\) ps) maintained a temperature of \SI{300}{K} and an isotropic pressure of 1.013 bar for 2 ns, thereby yielding a stress-free reference state (\(L_{0}\)) for subsequent deformation \cite{lee2021molecular}.\\\\
After equilibration, uniaxial tension was applied along the X-direction using an engineering strain rate of \(R_{\text{tensile}}=1\times10^{8}\,\text{s}^{-1}\). To preserve a true uniaxial state, lateral stresses were relaxed by coupling a Nosé--Hoover barostat to the Y and Z directions while the temperature was held at \SI{300}{K} \cite{chen2021simultaneously,gao2023molecular}. Engineering strain was computed as:
\[
\varepsilon(t)=\frac{L_{x}(t)-L_{0}}{L_{0}},
\]
and the axial and transverse Cauchy stresses were obtained from the virial components, \(\sigma_{ii}\). Stress--strain data were collected every 5 ps, whereas full atomic configurations---including centrosymmetry and per-atom potential energy---were dumped every 10 ps for post-processing. The total deformation simulation lasted for 2 ns, generating an engineering strain of 0.2.\\\\
The recorded trajectories were analysed to extract elastic moduli, yield strength and strain-hardening behaviour; dislocation activity and local structural changes were visualised using centrosymmetry and atomic energy fields. All quantitative comparisons between potentials (MEAM vs MTP) were performed on identically strained snapshots to isolate the effect of the force field.

\subsubsection{Solid-Liquid Phase Transition}
To probe the melting behaviour of the equiatomic CoCrFeMnNi high-entropy alloy, we performed direct‐heating simulations with both the MEAM potential of Choi \textit{et al.}\,\cite{choi2018understanding} and the moment-tensor potential (MTP) developed herein, using identical LAMMPS input decks. A cubic box of \(6\times6\times6\) FCC unit cells (\(864\) atoms, edge length \(6a_{0}=6\times3.518\;\text{\AA}=21.1\;\text{\AA}\)) was constructed under fully periodic boundary conditions. The as-built lattice was first relaxed at 0 K by conjugate‐gradient minimisation to energy and force tolerances of \(10^{-12}\,\mathrm{eV}\) and \(10^{-12}\,\mathrm{eV\,\AA}^{-1}\), respectively. Atomic velocities were then drawn from a Maxwell–Boltzmann distribution at \SI{300}{K}, and dynamics were integrated via the velocity–Verlet algorithm with a timestep \(\Delta t = 1\) fs \cite{PhysRevE.59.3733}.

Immediately following minimisation, no separate finite‐temperature equilibration was applied. Instead, the system was subjected to two successive isothermal–isobaric (NPT) ramps: a heating ramp from \SI{300}{K} to 2000 K over 3 ns, followed by a cooling ramp back to \SI{300}{K} over an additional 3 ns. Temperature and pressure were controlled by a Nosé–Hoover thermostat (\(\tau_{T}=0.1\) ps) and isotropic barostat (\(\tau_{P}=1\) ps) at 1.013 bar \cite{evans_nose_hoover_1985}. Thermodynamic data (temperature, volume, pressure, potential energy, enthalpy) were output every 1000 steps (1 ps), and full atomic configurations were dumped every 10 000 steps (10 ps) for structural analysis.

The resulting \(V(T)\) and \(H(T)\) traces—showing characteristic volume inflection and enthalpy plateaux—allowed us to bracket the alloy’s melting temperature and to directly compare the predictions of the Choi–MEAM and the present MTP parametrizations.  

\subsubsection{Nanoindentation}
Nanoindentation was modelled with LAMMPS using the MTP trained earlier, along with a separate run using MEAM potentials developed by Choi et al\cite{choi2018understanding}.  
The simulation used a cubic cell containing \(18\times18\times18\) FCC unit cells (23{,}328 atoms), identical to the model used for uniaxial tensile deformation. Prior to dynamics, the structure was first relaxed at 0\,K using conjugate-gradient minimization to remove high-energy overlaps. This was followed by thermal equilibration under the isothermal--isobaric (NPT) ensemble at 300\,K and 1.013\,bar for 2\,ns. Subsequently, a second minimization was carried out at 0\,K (energy and force tolerances of $10^{-4}$ eV and $10^{-5}$ eV $\AA^{-1}$ to eliminate residual stress introduced during equilibration, ensuring a well-relaxed initial state prior to indentation.

The simulation cell was periodic in X and Y and non-periodic (free) along Z.  
A \textit{mobile} group containing all atoms except a 30 \AA-thick bottom “anvil” slab (forces rigidly zeroed) was thermostatted at \SI{300}{K} with a Nosé–Hoover NVT scheme (damping time 50 fs). 

Indentation was effected with LAMMPS’ \texttt{fix indent} command: a rigid, frictionless spherical indenter of radius 30 \AA, centred at (32 \AA, 32 \AA, $z$), interacted with atoms through a repulsive harmonic potential
\[
F = k\,(R - r)\quad (r < R),
\]
where $k=16\;\mathrm{eV\,\AA}^{-3}$ and $R$ is the indenter radius, as used in previous nanoindentation works. \cite{alhafez2024nanoindentation,HAN20236027}.  
The indenter centre followed a piecewise linear trajectory
\[
z(t) = z_0 - v\,t \quad (0 \le t < t_{\mathrm{max}});\qquad
z(t) = z_0 - v\,(2\,t_{\mathrm{max}} - t) \quad (t_{\mathrm{max}} \le t \le 2\,t_{\mathrm{max}}),
\]
with $v=10\;\mathrm{m\,s}^{-1}$ and $t_{\mathrm{max}}=200\;\mathrm{ps}$.  
Thus, the full load–unload cycle spanned 400 ps (4×10$^5$ steps), reaching a maximum penetration depth of 20 \AA.

Atomic coordinates, forces, and per-atom stresses were saved every 5 ps in CFG format for post-processing, while the instantaneous indenter position and the three Cartesian components of the reaction force were logged every 0.1 ps.
\subsubsection{Energy-Volume Equation of State}
\subsection*{Static $E\!-\!V$ calculations for the Birch--Murnaghan fit}

To obtain the 0\,K equation of state (EOS) of equiatomic CoCrFeMnNi, we evaluated the static total energy of a primitive FCC cell (\(4\) atoms) over a broad range of lattice parameters and fitted the resulting \(E(V)\) data to the third‐order Birch–Murnaghan expression \cite{Murnaghan1944,Birch1947}.  
A uniform mesh of fourteen lattice constants from 3.30\,Å to 3.70\,Å (increment 0.02\,Å) was employed; the corresponding volumes span~\(\pm\!12\%\) about the experimental equilibrium value.

\paragraph*{DFT reference.}
Plane‐wave density‐functional calculations were performed in \textsc{VASP}\,\cite{kresse1996efficient} using the PBE exchange–correlation functional, projector‐augmented‐wave potentials, a kinetic‐energy cutoff of 500\,eV, and a \(15\times15\times15\) Monkhorst–Pack $k$‐mesh.  
Static single‐point energies were converged to $10^{-8}$ eV (EDIFF) with no ionic relaxation (NSW = 0, IBRION = -1, ISIF = 2).  
Smearing was treated with the Methfessel–Paxton scheme (ISMEAR = 1, SIGMA = 0.2 eV).  
The resulting energies serve as the benchmark for potential assessment.

\paragraph*{MEAM calculations.}
For each lattice constant the same 4‐atom cell was created in \textsc{LAMMPS}\,\cite{plimpton1995fast}, employing the equiatomic substitution sequence and atomic masses listed in the input deck.  
Interatomic interactions were described with the Choi \textit{et al.} MEAM parameter set. \cite{choi2018understanding}.  
The structure was relaxed at 0\,K via conjugate‐gradient minimisation (energy and force tolerances \(10^{-12}\,\mathrm{eV}\) and \(10^{-12}\,\mathrm{eV\,\AA}^{-1}\); 10\,000 iterations), after which the total energy \(E_{\mathrm{MEAM}}(V)\) was recorded.

\paragraph*{MTP calculations.}
Exactly the same set of cells was recomputed with our trained moment‐tensor potential.  
The input script performs an identical conjugate‐gradient minimisation to obtain the relaxed energy \(E_{\mathrm{MTP}}(V)\). The results are therefore directly comparable to MEAM and DFT.

\paragraph*{Fitting and derived parameters.}
For each method the set of \(\{E(V)\}\) points was fitted to the third‐order Birch--Murnaghan form  
\[
E(V)=E_{\min}+\frac{9B_{0}V_{0}}{16}\bigl[(\eta-1)^{3}B_{0}'+(\eta-1)^{2}(6-4\eta)\bigr],\qquad
\eta=(V_{0}/V)^{2/3},
\]
yielding the equilibrium volume \(V_{0}\), lattice constant \(a_{0}=(4V_{0})^{1/3}\) and bulk modulus \(B_{0}\).  
Equivalent BM‐EOS fitting strategies have proved reliable for CoCrFeMnNi and related HEAs \cite{ma2015Abinitio,Gao2024DefectEnergeticsHEA,shi2023first,Moreno2023ThermodynamicRHEA,Chen2020TiVNbMoHEA}.

\section{Results and Discussions}

We evaluated the reliability of the trained MTP against the well established MEAM potential across various physical properties prediction tasks such as Young's modulus, hardness and melting point by comparing the predicted values against experimental ones. The model was also tested against DFT calculations on structural properties such as equilibrium lattice constant,  energy, force and stress predictions on various test structures. These validations were performed to test the ability of trained MTP to capture and simulate physical and structural properties of CoCrFeMnNi over different size and scales. Further, we also performed and established strong scaling relationships for MEAM and MTP.\cite{mamun_comparing_2023} \cite{scheiber_ab_2016}

\subsection{Model Validation}

To validate the model against DFT on structural properties, we created two independent test sets, referred to as TS1 and TS2. TS1 consisted of 100 different SQS configurations sampled from Monte Carlo simulations using the software \emph{sqsgen} \cite{sqsgen} and TS2 consisted of 100 configurations generated on LAMMPS \cite{THOMPSON2022108171} sampled from a NVT trajectory performed for total of 1 ns after thermalisation for 200 ps to reduce correlation between the generated structures and the inital structure. The MD simulations were performed under 300 K using MEAM potential for CoCrFeMnNi. The structures were sampled uniformly every 10 ps to minimise correlation between structures within dataset and have an dataset independent from configurations on which the model was trained. For each of these structures we used the trained MTP force field to find the energy of the structures and forces per atom atoms in each structure. To obtain the DFT data for the same structures in TS1 and TS2, we performed Self consistent calculations for them on the VASP \cite{kresse1996efficient, kresse1996efficiency} package and obtained the energy and force per atom data for each structure by postprocessing results from the VASP standard output files.\\

\begin{figure}[H]
    \centering
\includegraphics[width=0.95\linewidth,keepaspectratio]{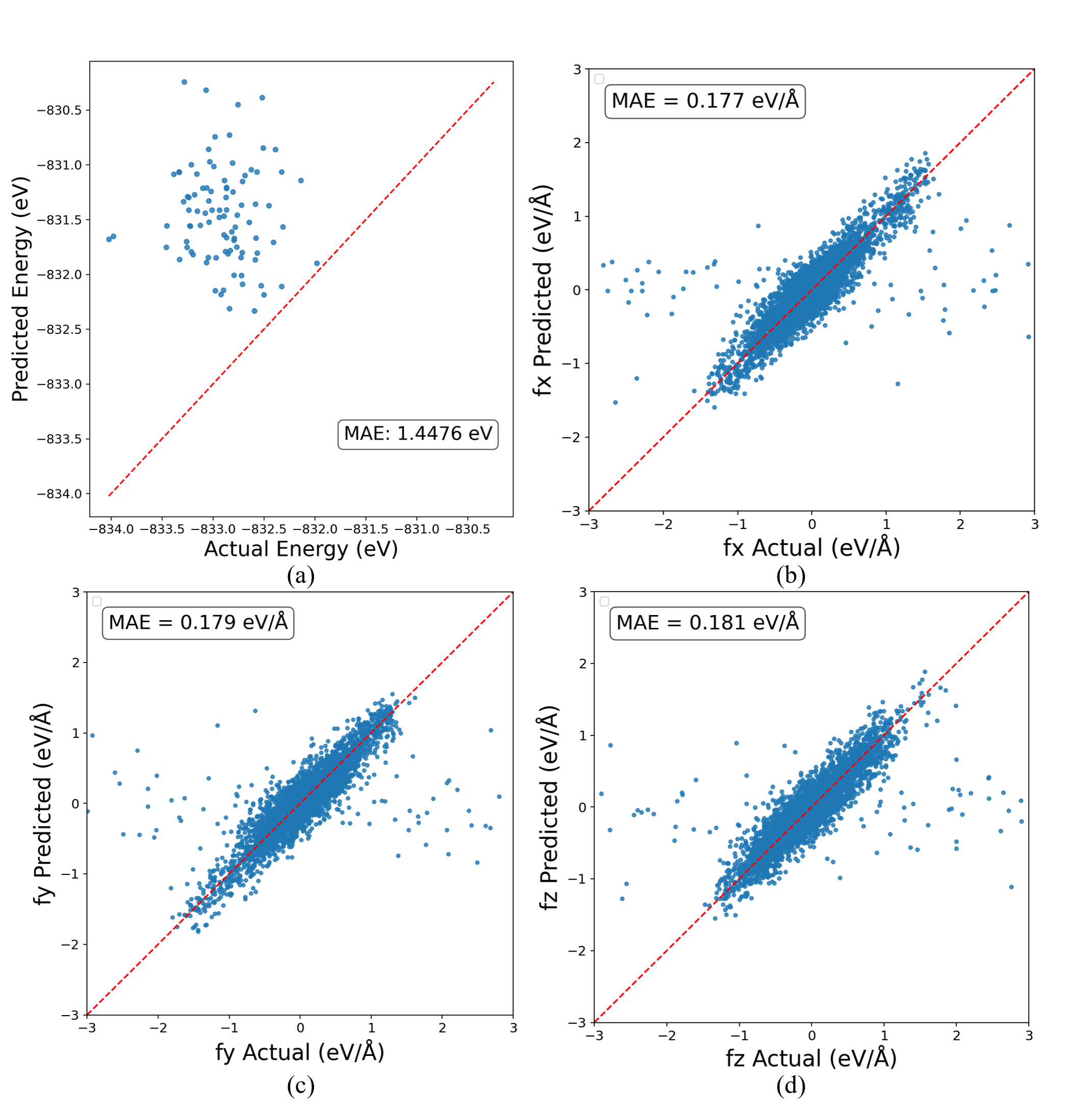}
\caption{Validation of the trained MTP on 100 distinct SQS structures. (a)predicted against actual energy in units of eV per structure in TS1. (b),(c),(d) predicted  against actual forces per atom in eV/Å for all the structures in TS1 along x,y and z axis respectively.}
    \label{fig:results_val_sqs}
\end{figure}

\begin{figure}[H]
    \centering
\includegraphics[width=0.95\linewidth,keepaspectratio]{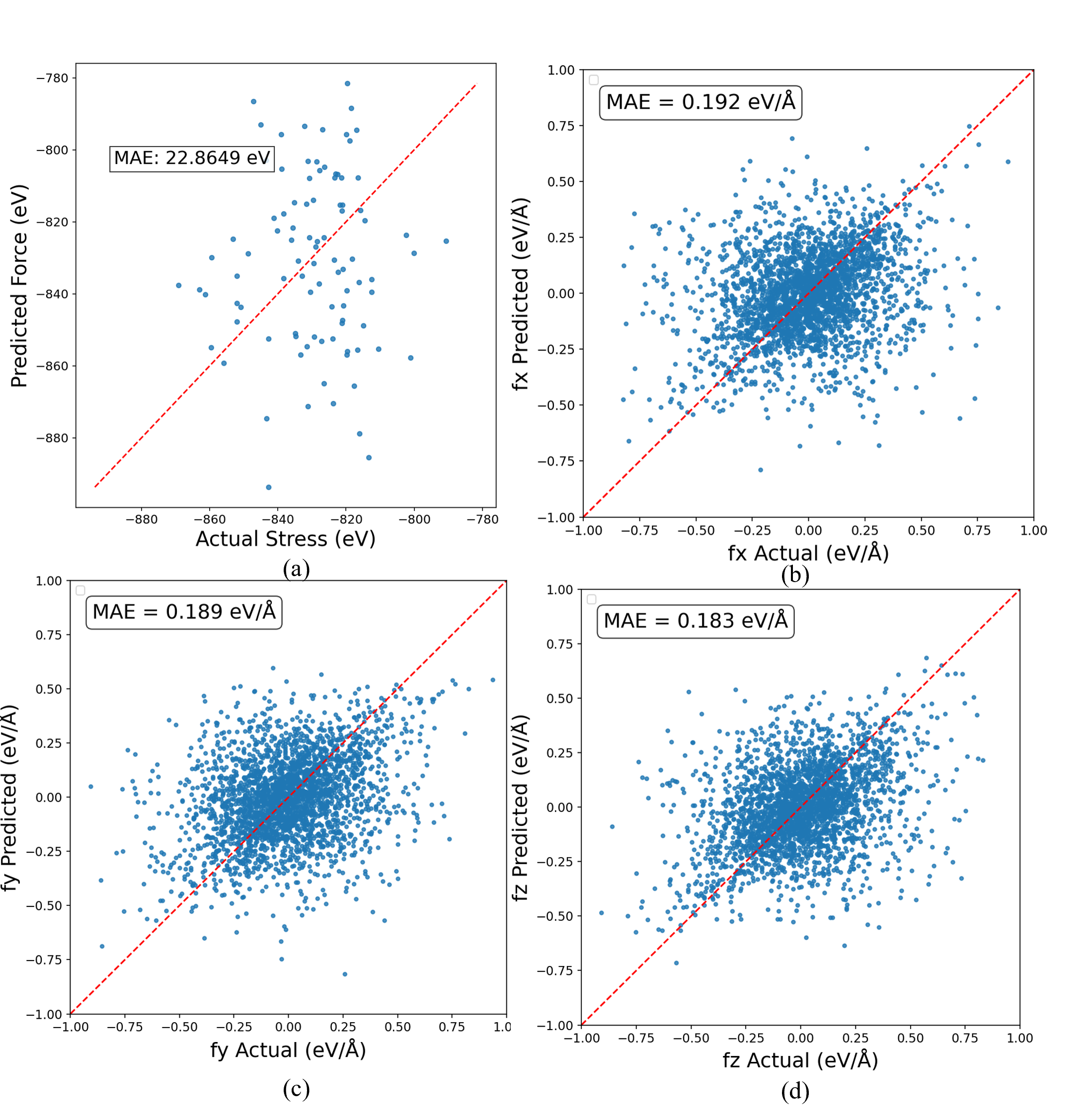}
    \caption{Validation of the trained MTP on 100 distinct structures sampled at equal intervals from NVT tajectory simulated on LAMMPS. (a) predicted against actual energy in units of eV per structure in TS2. (b),(c),(d) predicted  against actual forces per atom in eV/Å for all the structures in TS2 along x,y and z axis respectively.}
    \label{fig:results_val_lmp}
\end{figure}

Fig(\ref{fig:results_val_sqs}) and Fig(\ref{fig:results_val_lmp})
show the actual versus predicted energy and forces per atom for for TS1 and TS2 respectively. From the plots in Fig(\ref{fig:results_val_sqs}) it can be observed that the scatter points depicting ground state energy per configuration lie in a narrow band of $\approx 3 eV$. Although the trained MTP seems to be clearly overestimating the ground state energy, the MAE ($0.013 eV/atom$) for this data indicates that the relative errors in prediction is within reasonable limits and overall variation in energies between the structures is similar for the DFT (actual) and trained MTP (predicted) cases. This is consistent with the fact that all structures in TS1 are sampled from a single Metropolis Monte Carlo trajectory that meet the SQS criteria to a suboptimal grade and have a small variation between themselves. The trained MTP seems to reproduce relative energies of varied local environments with significant accuracy and overestimates the energy within a range of approximately $0.01$ $eV/atom$ for structures in TS1. The error metrics for force per atom predictions seems to be of similar accuracy along x,y and z axis at around 0.18 eV/Å. Some outliers can be seen in a region where the predicted force by the trained MTP seems to be small but the DFT predictions seems to be very large. Upon closer inspection, we found that all very poorly predcited forces can be traced to only 4 SQS structures out of the 100 total structures considered in the dataset, which maybe due to unphysical nature of their atomic configuration. As the structures are not relaxed, very large force predictions from DFT may suggest some large residual stresses in these structures, while the MTP regardless gives physically reasonable predictions as it is completely based on local descriptors of atomic environment and all the structures were created using equillibrium structure parameters.  \\

\begin{table}[H]
  \centering
  \caption{Mean Absolute Error (MAE) of energy and force/atom predictions for the  TS1 test set.}
  \label{tab:mtp_errors}
  \begin{tabular}{lccc}
    \toprule
    Quantity & MAE\\
    \midrule
    Energy & $0.013\ \mathrm{eV/atom}$ \\
    Force X & $0.177\ \mathrm{eV/\AA}$ & \\
    Force Y & $0.179\ \mathrm{eV/\AA}$ &\\
    Force Z & $0.181\ \mathrm{eV/\AA}$ & \\
    \bottomrule
  \end{tabular}
\end{table}

Comparing the predictions of our model on TS1 with predictions in previous literature, we found that these values are very similar to that obtained from other MTPs reported by Jafary-Zadeh et al. \cite{jafary2019applying}, who obtained a MAE for force of $\approx0.4 eV/\AA$ when modeling CoFeNi using a similar approach, and Pandey et al.\cite{pandey2022machine} who achieved energy MAEs of $\approx0.08 eV/atom$ for quaternary MoNbTaW HEAs. This data shows near \emph{ab initio} accuracy as expected from the near equilibrium configurations present in TS1.

We then performed a similar set of validations for structures in the TS2 set. We observed that the MAE in energy predicted ($0.211$ $eV/atom$) in this case is much larger than in TS1. But it also relieving to note that we do not observe any systematic bias in energy prediction unlike TS1; this suggests that the slight systematic over-prediction of energy in TS1 maybe limited to the generated non relaxed SQS configurations from TS1 only. From the force prediction against actual values plots we again observe that there seems to be large scatter around the ideal fit line although the Mean Absolute Error (MAE) for neither of the axes is very large. These results are somewhat expected as we tried to make the test dataset as independent as possible and it can contain many configurations which maybe considered very extrapolative with respect to the data on which the model is trained. This large error is primarly observed as the compositon of the configurations in TS2 is very different from the actual dataset on which the MTP model was trained. 

In the case of TS2, despite energies spanning a wider range of roughly $–880eV$ to $–780eV$, which is expected due to the large variation in structures obtained from the NVT trajectory, the MTP captures the overall energy fluctuations without bias; the error in forces ($0.19$ $eV/\AA$) remains close to the ideal fit, showing that the force field correctly quantifies direction of atomic forces under finite perturbations. Despite the large errors in energy for TS2 configurations due to considerably different composition of compared to actual training set used. The general errors in the predictions of energy and force per atom from our MTP model are in accordance with previous studies despite the increased chemical complexity and a significantly smaller training data set for the five component high entropy alloy \cite{han_accuracy_2023} \cite{rosenbrock2021machine}. This validates the physicality of the trained MTP.

\begin{table}[H]
  \centering
  \caption{Mean Absolute Error (MAE) of energy and force/atom predictions made on TS2}
  \label{tab:md_errors}
  \begin{tabular}{lccc}
    \toprule
    Quantity & MAE \\
    \midrule
    Energy & $0.211\ \mathrm{eV/atom}$  \\
    Force X & $0.192\ \mathrm{eV/\AA}$ \\
    Force Y & $0.189\ \mathrm{eV/\AA}$ \\
    Force Z & $0.183\ \mathrm{eV/\AA}$ \\
    \bottomrule
  \end{tabular}
\end{table}

\subsection{Vacancy Formation Energy}

To further validate the MTP in predicting defect energetics, we compared the vacancy formation energies obtained from the MTP against DFT calculations and values derived from the MEAM potential. To minimize finite-size effects, local atomic environment calculations were carried out using both DFT and the trained MTP on a 108-atom SQS structure. Three different vacancy-induced structures were prepared for each element. We followed a similar methodology to that used to calculate the vacancy formation energy in FeMnCoCrNi by Gao \textit{et al.}~\cite{gao2024defect}.

Initially, the chemical potential for each element was calculated, and the average vacancy formation energy—regardless of atomic type—was obtained from two of the three structures prepared for each atom (10 in total). The chemical potential for each element type \(X\) is defined as follows:

\begin{equation}
\mu_X = E_0 + \left\langle E_V^{f} \right\rangle 
- \frac{1}{N_X} \sum_{k=1}^{N_X} E_{d,k},
\end{equation}

where \(E_0\) is the total energy of the defect-free structure, \(E_{d,k}\) is the total energy of the structure containing a vacancy of element type \(X\) at site \(k\), and \(N_X\) is the number of vacancy structures of element \(X\).

The average vacancy formation energy is then computed as:

\begin{equation}
\left\langle E_V^{f} \right\rangle 
= \frac{1}{N_T} \sum_{n=1}^{N_T} E_{d,n}
- \frac{N-1}{N} E_0,
\end{equation}

where \(N_T\) is the total number of single-vacancy structures considered, and \(N = 108\) is the number of atoms in the SQS supercell.

Finally, another vacancy-induced structure was used to compute the vacancy formation energy using:

\begin{equation}
E_{f} = E_{d} - E_0 + \mu_X,
\end{equation}

where \(E_{f}\) is the vacancy formation energy of element \(X\), and \(E_{d}\) is the total energy of the structure containing a vacancy of type \(X\) at a site not previously used to calculate the average vacancy formation energy \(\left\langle E_V^{f} \right\rangle\).

The values obtained for each constituent element are reported in Table~\ref{tab:vac}. While the DFT and MTP predictions do not perfectly match for individual elements, the comparison of average vacancy formation energies provides a more meaningful benchmark. The average vacancy formation energy from DFT is \(E_{\mathrm{vac}}^{\mathrm{DFT}} = 2.342~\text{eV}\), from MEAM is \(E_{\mathrm{vac}}^{\mathrm{MEAM}} = 3.226~\text{eV}\), and the MTP predicts \(E_{\mathrm{vac}}^{\mathrm{MTP}} = 2.003~\text{eV}\). The average values from MTP and DFT are in closer agreement with each other compared to the MEAM result. The relatively high energies are expected, as removing a single atom from a 108-atom supercell corresponds to a vacancy concentration of approximately \(c_{\mathrm{vac}} \approx 0.9\%\), which in equilibrium represents very high-temperature conditions.

\begin{table}[H]
\centering
\caption{Comparison of vacancy formation energies predicted by the trained machine-learned interatomic potential (MTP) and parameterized (MEAM) potential against those obtained from DFT calculations for each atomic species. Values are reported in eV.}
\label{tab:vac}
\begin{tabular}{l S[table-format=1.3] S[table-format=1.3] S[table-format=1.3]}
\toprule
\textbf{Element} & {\textbf{$E_{\text{MEAM}}$ (\si{\electronvolt})}} & {\textbf{$E_{\text{MTP}}$ (\si{\electronvolt})}} & {\textbf{$E_{\text{DFT}}$ (\si{\electronvolt})}} \\
\midrule
Fe & 3.482 & 1.913 & 2.301 \\
Mn & 3.094 & 2.094 & 2.305 \\
Co & 3.415 & 1.801 & 2.600 \\
Cr & 2.930 & 2.001 & 2.357 \\
Ni & 3.161 & 2.308 & 2.255 \\
\bottomrule
\end{tabular}
\end{table}

\subsection{Lattice parameter and Bulk modulus}
\begin{figure}[H]
    \centering    \includegraphics[width=0.7\linewidth,keepaspectratio]{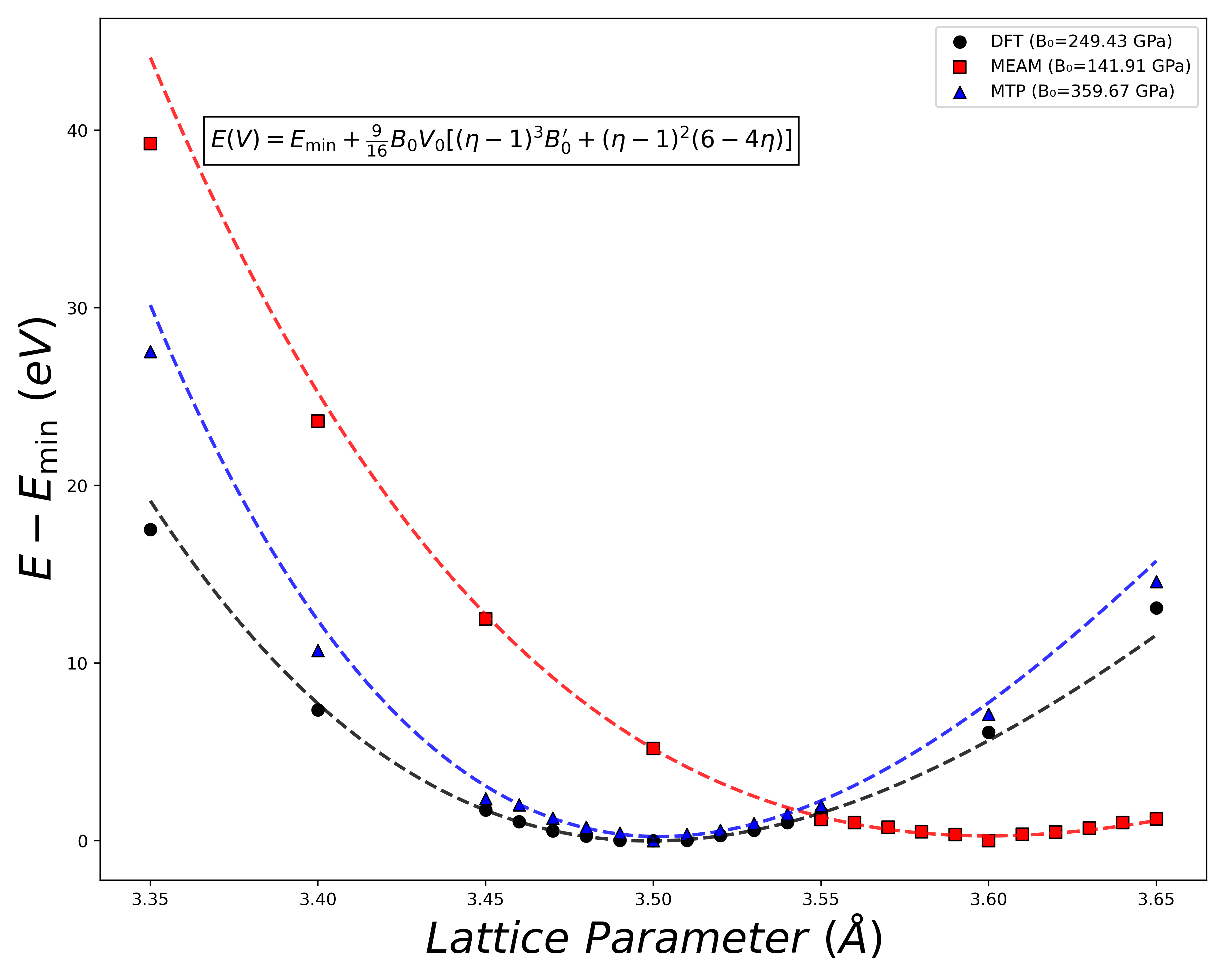}
\caption{
       {DFT, MEAM and MLIP Energy - Volume EOS comparisons for the HEA.}
        Energy deviations from the minimum energy configuration ($E - E_{min}$) are plotted as a function of the lattice parameter. Dashed curves represent Birch Murnaghan fits to DFT, MEAM and our trained MLIP results.
    }
    \label{fig:EOS_results}
\end{figure}

The accuracy of trained MTP against well established MEAM and DFT in predicting the equilibrium lattice parameter and the energy versus volume equation of state (EOS) was undertaken and is shown in Fig(\ref{fig:EOS_results}). The variation in energy with respect to the lattice parameter is depicted after subtraction of the minimum (reference) energy for each curve to allow for a one-to-one comparison.

Initially, the validation of the developed MTP was performed by calculating the lattice constant ($a_{0}$) and bulk modulus ($B_0$). It was achieved by fitting the Birch-Murnaghan equation of state (BM-EOS) of third order \cite{Murnaghan1944, Birch1947} to the curve of the energy versus lattice parameter, as shown in Fig(\ref{fig:EOS_results}). It represents the corresponding EOS for a range of lattice parameters close to the predicted equilibrium value for each of the potentials.

\begin{equation}
    E(V) = E_{min} + \frac{9B_0V_0}{16}[(\eta-1)^3B_0'+(\eta-1)^2(6-4\eta)]
\end{equation}
where
\begin{equation}
\eta=(\frac{V_0}{V})^{2/3} 
\end{equation}

In the above equation, $V_0$, $E_{min}$, V and $B_0$ corresponds to the equilibrium volume, corresponding energy, volume of change and the bulk modulus, respectively. However, $B_0'$ is a constant. The above equation has been used successfully to estimate the lattice parameter and bulk modulus in CoCrFeMnNi HEA \cite{ma2015Abinitio}\cite{Gao2024DefectEnergeticsHEA} and other HEA families \cite{shi2023first,Moreno2023ThermodynamicRHEA,Chen2020TiVNbMoHEA}.
It can be clearly observed that the location of minima for DFT and trained MTP almost exactly match at 3.50 \AA, whereas the minima for MEAM is located at around 3.6 \AA.

The fitted values of $B_{0}$ obtained from MTP, MEAM and DFT are 359.67 GPa, 141.91 GPa, and 249.43 GPa, respectively. Previous experimental and first principle calculations also predict values between range of 135 GPa to 155 GPa \cite{Ma2015_ActaMater}\cite{Yu2016_JApplPhys}\cite{Zhang2017_NatCommun}. Though the predicted $B_{0}$ for the MTP deviates from the DFT predictions (by $\approx$ 44 $\%$), we note that the equilibrium lattice parameter is faithfully reproduced by the MTP. Further improvement in the prediction of the modulus could be attained with a more extensive training data set of configurations away from equilibrium.

\subsection{Tensile deformation}
                        
\begin{figure}[H]
    \centering
    \includegraphics[width=0.7\linewidth,keepaspectratio]{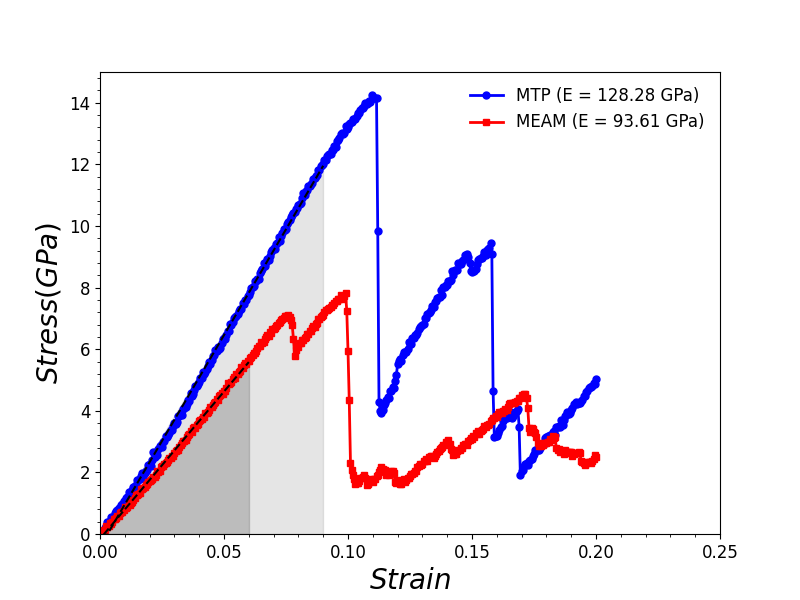}
    \caption{Comparison of the uniaxial tensile deformation between the trained MTP and MEAM in blue and red curves respectively. The Young's modulus indicated in the plot is extracted from the linear portion of the curve, under elastic region indicated in grey.}
    \label{fig:stressstrainplot}
\end{figure}

Uniaxial extension simulations were performed on a 18$\times$18$\times$18 supercell of CoCrFeMnNi for the trained MTP and MEAM, to ascertain the mechanical properties. The linear portion of the stress-strain curve, shown as the shaded region (corresponding to a strain $\approx 0–0.08$) in Fig(\ref{fig:stressstrainplot}) was used to extract the Young's modulus. The MTP yields E=128 GPa, closer to DFT benchmarks than the softer MEAM estimate of E=94 GPa. We can see that beyond the shaded bands (strain $\approx 0–0.08$) the response departs from linearity, signalling the onset of plasticity. Beyond the elastic limit both curves exhibit serrated yielding sharp stress drops followed by reloading as discrete dislocation bursts or local phase changes relieve internal stress before further hardening. The difference in maximum stress and limiting strain for MTP and MEAM can be explained in terms of the overall density and mobility of dislocations produced under each of these interatomic potentials. Dislocations start appearing much earlier for MEAM and seem to undergo much faster motion and multiple recombinations as can be observed from the more serrated behavior of tensile stress-strain curve.

\subsection{Hardness measurements}

\begin{figure}[H]
    \centering
\includegraphics[width=0.7\linewidth,keepaspectratio]{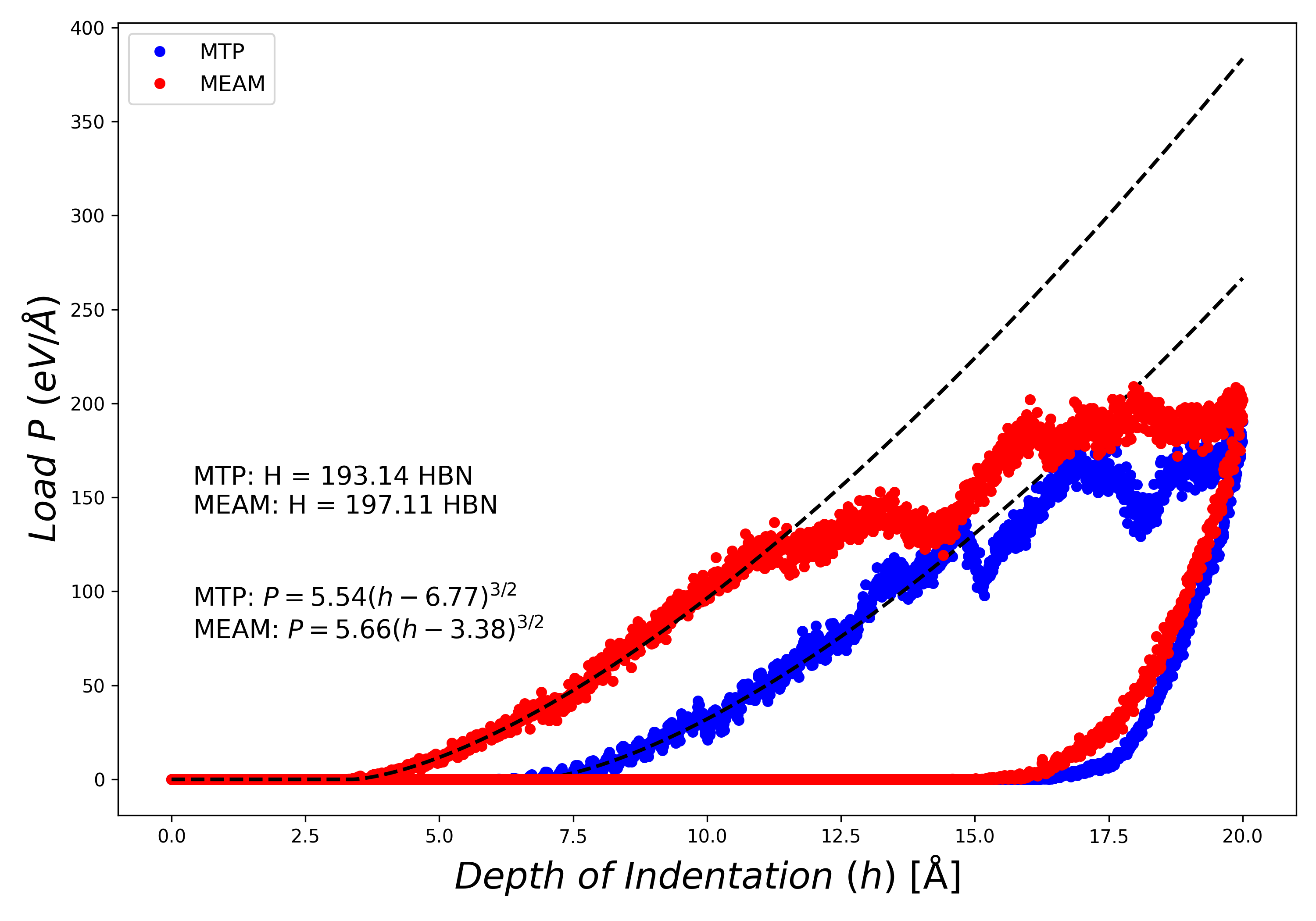}
   \caption{Comparison of nanoindentation simulations for trained MTP and MEAM in blue and red respectively. The fitted P versus h equation and Hardness on Brinell scale predicted by each of the potentials are indicated.}
  \label{fig:results_nanoindent}
\end{figure}

We calculated the hardness of the Cantor alloy through nanoindentation simulations performed on a 18$\times$18$\times$18 supercell (23,328 atoms) of CoCrFeMnNi, modeled using our MTP and compared it to the results from the MEAM potential. A spherical indenter with radius $R = 30\,\text{\AA}$ was used. The fitted curve follows the relationship $P = \beta\, (h-h_o)^{3/2}$, where $P$ is the load-dependent response and $h-h_o$ is the depth of the indentation. The fitted parameters and the calculated hardness values on the Brinell scale are presented in  Fig(\ref{fig:results_nanoindent}).\\

The Brinell hardness extracted from our nanoindentation simulations, \(H_{\rm MTP}=193.1\) HB, lies in the range 150 to 200 HB range reported experimentally for annealed CoCrFeMnNi alloys \cite{Stepanov2015}. So, our MTP prediction matches the experiment within 3\%.\\

\subsection{Solid-Liquid Phase Transition}

\begin{figure}[H]
    \centering
    % Include the PNG image with high-quality rendering
    \includegraphics[width=0.7\linewidth,keepaspectratio]{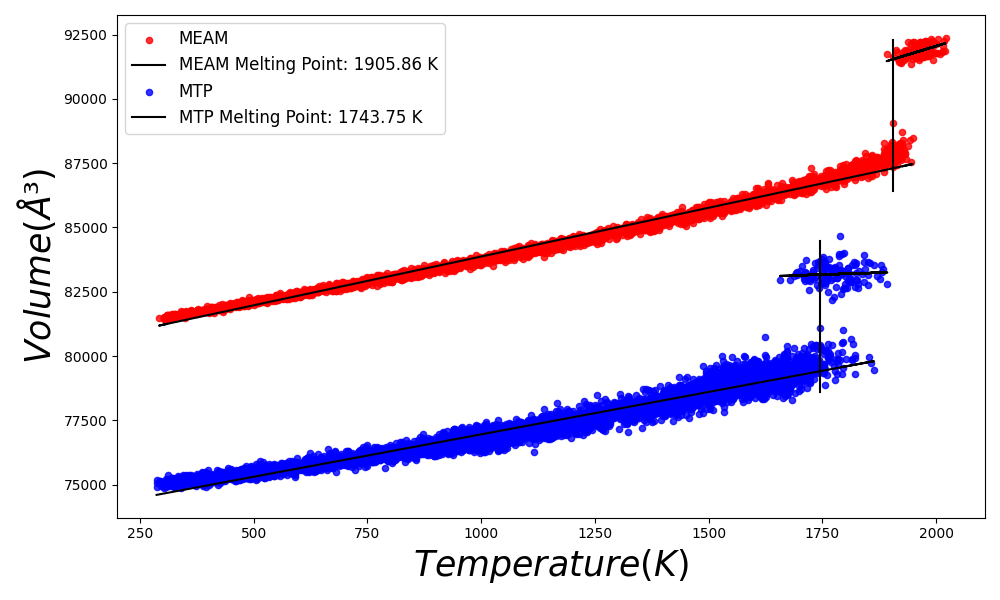}    
    \caption{Comparison of heating curves for trained MTP and MEAM in blue and red respectively. The discontinuous jump in volume against temperature as observed in the plots indicated the first order solid-liquid phase transition.}
    \label{fig:results_phase_transition}
\end{figure}

The comparison of solid-liquid phase transition was performed on an equilibrated supercell of CoCrFeMnNi with a heating rate of $r = 350 K/ns$ from $300 K$ to $2000 K$, conducted with MEAM and trained MTP\cite{asadi2015quantitative}. The volume jump shows the phase transition point. The phase transition points as indicated in Fig(\ref{fig:results_phase_transition}) appear at 1743.8 K for MTP and around 1905.9 K for MEAM. The MTP prediction matches much more closely to the experimentally reported value of 1557 K\cite{Werner_Skrotzki2023MT-MF2022050} for Co$_{20}$Cr$_{26}$Mn$_{20}$Fe$_{20}$Ni$_{14}$. But as reported in previous previous MD\cite{HAAPALEHTO2022111356} studies and through well established theory, the kinetics of the temperature increase process can significantly overestimate the actual melting point temperature. To compensate for the effect of heating rate on the predicted melting temperature ($T_{M}$) , we also estimated the $T_{M}$ using $50\%$-$50\%$ solid-liquid coexistence simulations. Where the  $T_{M}$ is defined as the temperature point at which equal volume of solid and liquid phases co-exist in equilibrium with each other. For the trained MTP and MEAM we achieve a $50\%$-$50\%$ solid-liquid composition at 1600 K and 1800 K respectively. Although structures modeled with the trained MTP are unstable beyond temperatures as high as 1800 K, its ability to predict  $T_{M}$ accurately demonstrates the extrapolative nature of MTP as no liquid phase data samples were included in the inital training set. Further, while the current version of the trained model may only be capable of modelling lower temperature dynamics, it can be easily trained for applications at higher temperature or other extreme conditions with minimal additional training rather than starting from a new instance of MTP.

\subsection{Scaling Performance}

\begin{figure}[H]
    \centering
    % Include the PNG image with high-quality rendering
    \includegraphics[width=0.5\linewidth,keepaspectratio]{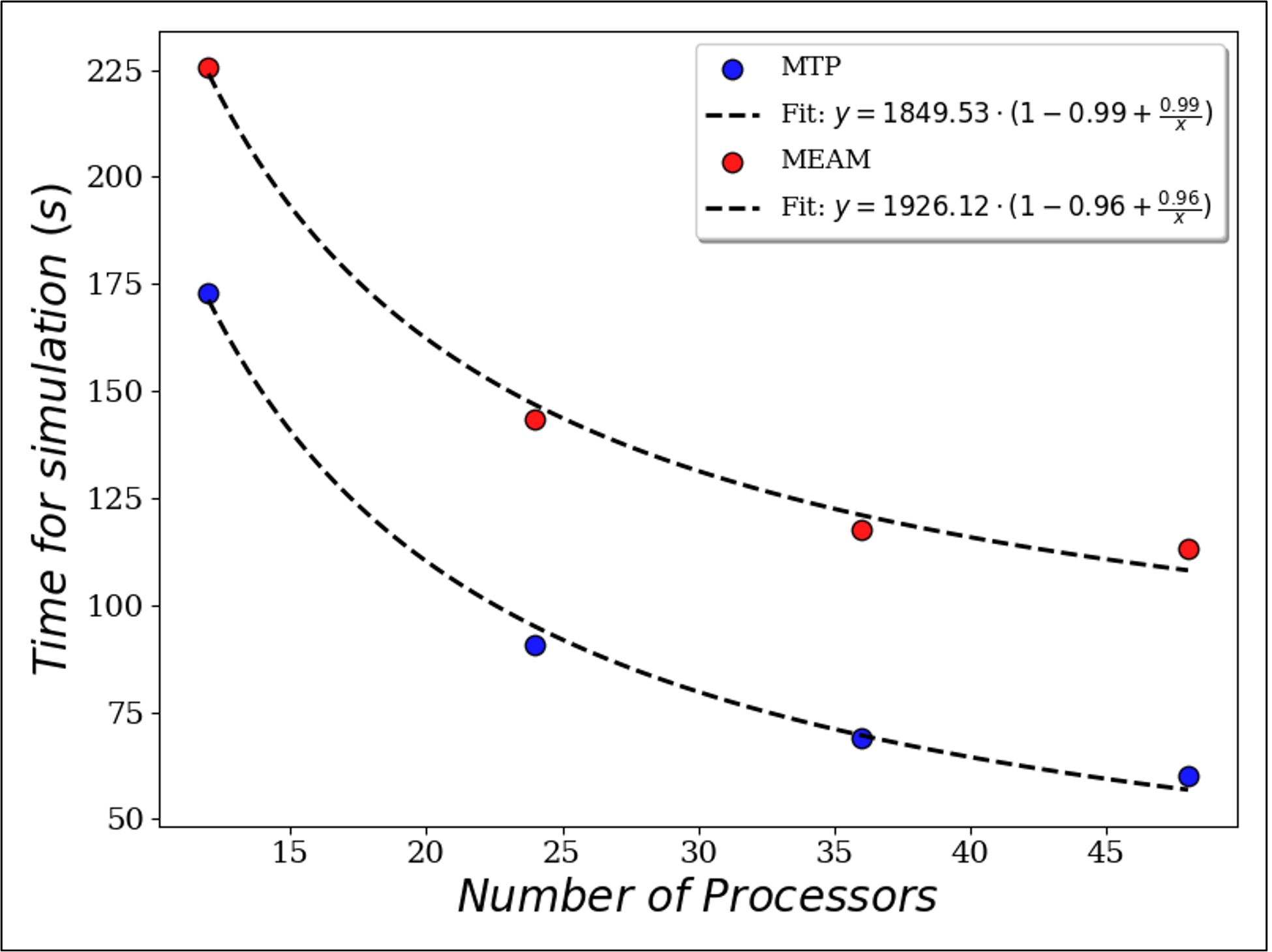}
\caption{Strong‐Scaling performance comparision of MTP and MEAM shown by the blue and red markers respectively. Amdahl's scaling law fitted to computation time for MTP and MEAM data in black dashed lines.}
    \label{fig:results_scaling}
\end{figure}

NVT simulations for 6x6x6 HEA system at 300 K performed on LAMMPS exhibit distinct scaling characteristics for MTP and MEAM potentials as processor counts increase. Initial runs at 12 MPI ranks show MTP's advantage, completing 10K steps in 173 seconds versus MEAM's 225 seconds. When scaling to 24, 36, and 48 processors, MTP's runtime decreases substantially to 91s, 69s, and 60s respectively, significantly faster than MEAM (see Fig(\ref{fig:results_scaling})). This widening performance gap follows Amdahl's scaling law:
$$T(N) = T_{\infty} \left(1 - p + \frac{p}{N}\right)$$

where $p$ denotes the parallelizable fraction. MTP's exceptional $p=0.99$ indicates near-total parallelization, outperforming MEAM's $p=0.96$. The minimal serial component in MTP enables near linear speedup at moderate processor counts.\\

The MTP's design delivers superior efficiency on high performance systems, retaining 35\% parallel efficiency at 48 processors compared to MEAM's 24\%. This 46\% relative improvement stems from MTP's localized descriptor calculations and reduced communication overhead\cite{nguyen2021billion}. The efficiency gains confirm MTP's viability for large-scale alloy studies, providing DFT level accuracy without compromising performance particularly crucial for exascale architectures where communication minimization determines practical applicability.

\section{Conclusions}

We developed a machine–learned interatomic potential (MLIP) for the equiatomic CoCrFeMnNi high–entropy alloy (Cantor alloy) using a training corpus that combines SQS disorder, defect diversity, and active, on-the-fly data acquisition. Starting from an SQS supercell that approximates an ideal equiatomic ratio within the 108-atom cell (generated with \texttt{mcsqs}/ATAT\cite{atat2}), we augmented the dataset with targeted defects vacancies (three atoms removed from the bulk, yielding a 105-atom cell under PBC), dislocations (removal of a half-plane along $\langle 111\rangle$ placed off-center), and stacking faults (one-third interplanar shift perpendicular to $\langle 111\rangle$) at \SI{300} {K} along with with AIMD frames across temperatures of 600 K and 900K computed in \textsc{VASP}\cite{kresse1996efficient}. We evaluated several Moment Tensor Potential (MTP) architectures\cite{mtp,novikov2020mlip} with differing levels and radial basis sizes (L8–RB8, L8–RB10, L16–RB8; cf.\ Table~\ref{tab:mtp_tradeoffs}). Guided by performance on validation errors and stability, we selected L8–RB10 as the final model and retrained it on the base + active-learned data. Active learning followed the MaxVol/extrapolation-grade paradigm\cite{gubaev2019accelerating} implemented in MLIP, interfaced to LAMMPS\cite{plimpton1995fast}: MD trajectories were screened for high extrapolation grade, diverse high-$\gamma$ snapshots were DFT-labeled, appended to the set, and the MTP was retrained until no further excursions occurred.\\\\
Head-to-head comparisons with a MEAM parameterization\cite{choi2018understanding} show that the MTP reproduces the equation of state more faithfully, yielding an equilibrium lattice parameter in near-quantitative agreement with DFT results. While estimating the vacancy formation energy, the MTP estimates on average are significantly closer to the DFT predictions compared with MEAM; under uniaxial tension, the MTP captures the elastic modulus and the onset of plasticity with characteristic serrated flow; in nanoindentation, the hardness falls within the experimentally reported range for annealed CoCrFeMnNi. Besides, MTP also predicts a more realistic melting point in comparison to the MEAM potential. These trends are consistent with the model-selection results in Table~\ref{tab:mtp_tradeoffs}, where modest radial enrichment (RB\,=\,10) at Level~8 MTP improved fidelity and MD robustness without the steep cost of Level~16.\\\\
A practical advantage of the present MTP is the numerical efficiency. Strong-scaling tests Fig(\ref{fig:results_scaling}) show shorter wall times, approximately 2 to 3 times faster simulations and higher parallel efficiency than MEAM at moderate core counts for any reasonable High Performance Computing facilities, making the model an effective potential for large-scale simulatiaons that approach classical-MD throughput while retaining near-DFT fidelity for the trained thermodynamic and structural window.\\\\
While energies and forces are reproduced well; the bulk modulus is overestimated despite improved trends in comparison to MEAM, besides, the transferability degrades when sampling far beyond the trained ranges of physical conditions such as temperature, strain and lattice parameters. Magnetism and short-range order were not treated explicitly in the present reference data. The workflow is, however, intrinsically extensible: the same active-learning loop can fine-tune the potential to new conditions (e.g., higher temperatures, liquid configurations, CSRO-rich motifs, or spin-polarized DFT) by adding targeted reference frames and retraining.\\\\
To support reuse and benchmarking, we provide: (i) scripts for the generation of all AIMD datasets via GitHub (including SQS, defect builders, and AL drivers), and (ii) LAMMPS-ready potential files for open-source use. We expect MLIP dataset generation and model training procedure to transfer readily to non-equiatomic and second-generation HEAs with minimal adaptation of the sampling schedule and hyperparameters.

\section{Acknowledgement}
Financial assistance received from the Science and Engineering Research Board (SERB), India (under project SRG/2020/002449) is acknowledged. The authors also acknowledge the use of the Param Ananta and Param Porul supercomputing facilities to carry out the simulations and training.

\section{Data Availability}
All DFT structures generated using VASP \cite{kresse1996efficient,kresse1996efficiency}, the final MTP developed for the CoCrFeMnNi alloy using the MLIP Package \cite{novikov2020mlip}, MTP parameter files, results and analysis scripts used in this work are openly available at \url{https://github.com/cocokane/CantorAlloyMLIP}. The repository is written to enable full reproduction of our workflow and serve as a reference for training a custom MTP with active learning using the MLIP package. 

\bibliography{references.bib}
\end{document}